\documentclass[12pt]{article}
\usepackage{graphics,graphicx,graphpap,color}
\usepackage{setspace}

\newcommand\bmat{\left( \begin{array}{cc}}
\newcommand\emat{\end{array}\right)}

\def\msbar{\ifmmode{\overline{\rm MS}} \else{$\overline{\rm MS}$} \fi}
\def\drbar{\ifmmode{\overline{\rm DR}} \else{$\overline{\rm DR}$} \fi}

\def\ti              {\tilde}

\def\a               {\alpha}
\def\b               {\beta}
\def\d               {\delta}
\def\D               {\Delta}

\def\g               {\gamma}
\def\G               {\Gamma}
\def\l               {\lambda}
\def\t               {\theta}

\def\x               {\chi}

\def\sf              {{\ti f}}

\def\sfL             {{\ti f_L^{}}}
\def\sfR             {{\ti f_R^{}}}

\def\st              {{\ti t}}
\def\sb              {{\ti b}}
\def\stau            {{\ti \tau}}

\def\chp             {\ti \x^+}

\def\nt              {\ti \x^0}

\newcommand{\msf}[1]   {m_{\ti f_{#1}}}

\def\tw              {\t_{\scriptscriptstyle W}}

\def\vor             {\frac{1}{(4\pi)^2}}
\def\kslash          {\not{\!k}\,}

\def\non             {\nonumber}

\renewcommand\d{\delta}

\begin{document}

\pagestyle{empty} \vspace*{-1cm}

\begin{flushright}
  hep-ph/0308146
\end{flushright}

\vspace*{2cm}

\begin{center}
{\Large\bf\boldmath
   Improved full one--loop corrections 
   \\[2mm]
   to $A^0 \rightarrow \sf_1 \,\ {\bar{\!\!\sf}}_{\!\!2}$
   and {$\sf_2 \rightarrow \sf_1 A^0$}} \\[5mm]

\vspace{10mm}

C.~Weber, H.~Eberl, W.~Majerotto\\[5mm]

\vspace{6mm}
{\it Institut f\"ur Hochenergiephysik der \"Osterreichischen
Akademie der Wissenschaften, A--1050 Vienna, Austria}

\end{center}

\vspace{20mm}

\begin{abstract}
We calculate the {\em full} electroweak one--loop corrections to 
the decay of the CP--odd Higgs boson $A^0$ into scalar fermions in 
the minimal supersymmetric extension of the Standard Model. For 
this purpose many parameters of the MSSM have to be properly 
renormalized in the on--shell renormalization scheme. We have also 
included the SUSY--QCD corrections. For the decay into bottom 
squarks and tau sleptons, especially for large $\tan\b$, the 
corrections can be very large making the perturbation expansion 
unreliable. We solve this problem by an appropriate definition of 
the tree--level coupling in terms of running fermion masses and 
running trilinear couplings $A_f$. We also discuss the decay of 
heavy scalar fermions into light scalar fermions and $A^0$. We 
find that the corrections can be sizeable and therefore cannot be 
neglected. 
\end{abstract}

\vfill
\newpage
\pagestyle{plain} \setcounter{page}{2}

\section{Introduction}
The search for a Higgs boson is the primary goal of all present 
and future high energy experiments at the TEVATRON, LHC or an $e^+ 
e^-$ Linear Collider. Whereas the Standard Model (SM) predicts 
just one Higgs boson, with the present lower bound of its mass 
$m_H \geq 114.4~\mbox{GeV}$ (at 95\% confidence level) 
\cite{CERN-EP}, extensions of the SM allow for more Higgs bosons. 
In particular, the Minimal Supersymmetric Standard Model (MSSM) 
contains five physical Higgs bosons: two neutral CP--even ($h^0$ 
and $H^0$), one neutral CP--odd ($A^0$), and two charged ones 
($H^\pm$) \cite{GunionHaber2, GunionHaber1}. The existence of a 
charged Higgs boson or a CP--odd neutral one would give clear 
evidence for physics beyond the SM. For a discovery precise 
predictions for its decay modes and their branching ratios are 
necessary. In case supersymmetric (SUSY) particles are not too 
heavy, the Higgs bosons can also decay into SUSY particles 
(neutralinos $\nt_i$, charginos $\ti \x^\pm_k$, sfermions 
$\sf_m$), $H^0, A^0 \rightarrow \nt_i \nt_j$ ($i,j$ = 1\ldots 4), 
$H^0, A^0 \rightarrow \ti \x^+_k \ti \x^-_l$ \mbox{($k,l$ = 1, 
2)}, $H^0, A^0 \rightarrow \sf_m \,\bar{\sf_n}$ ($m,n$ = 1, 2), 
$H^\pm \rightarrow \ti\x^+_k \nt_i$, $H^\pm \rightarrow \sf_m 
\,\bar{\sf'_n}$. At tree--level, these decays were studied in 
\cite{tree1, tree2}. In particular, the branching ratios for the 
decays into sfermions, $H^0, A^0 \rightarrow \sf_m \,\bar{\sf_n}$ 
, can be sizeable depending on the parameter space \cite{Bartl1, 
Bartl2}. The SUSY--QCD corrections to the decays into sfermions 
have also been calculated \cite{SUSY-QCD, Arhrib}. The corrections 
to the decays into neutralinos \cite{0111303} or charginos 
\cite{Zhang} due to fermion/sfermion exchanges have also been 
found non--negligible. 
\\[3mm]
\noindent In this paper, we study in detail the decay of the 
CP--odd Higgs boson $A^0$ into two sfermions, $A^0 \rightarrow 
\sf_1 \,\bar{\sf_2}$. In particular, the third generation 
sfermions $\st_i, \sb_i$, and $\stau_i$ are interesting because 
one expects them to be lighter than the other sfermions due to 
their large Yukawa couplings and left--right mixings. Since $A^0$ 
only couples to $\sf_L$--$\sf_R$ (left--right states of $\sf$), 
and due to the CP nature of $A^0$, $A^0 \rightarrow \sf_i \ 
\bar{\!\sf_i}$ vanishes. (This is valid also beyond tree--level 
for real parameters in the MSSM.) We will calculate the {\em full} 
electroweak corrections in the on--shell scheme. Owing to the fact 
that almost all parameters of the MSSM have to be renormalized in 
this process and hence a large number of graphs has to be 
computed, the calculation is very complex. Despite this 
complexity, we have performed the calculation in an analytic way. 
We have also studied the crossed channel $\sf_2 \rightarrow \sf_1 
A^0$. Some important numerical results, especially for the decays 
of $A^0$ into the third generation squarks and the corresponding 
crossed channels have already been shown in \cite{0305250}. 
\\[3mm]
\noindent This paper has three new elements. First, we give in the 
Appendix all analytical formulae needed for the calculation. 
Second, we describe in detail our method for improving the full 
one--loop calculation. As pointed out in \cite{0305250}, in the 
case of the decay of $A^0 \rightarrow \sb_1 \!\bar{\ \sb_2}$ or 
$A^0 \rightarrow \stau_1 \bar{\,\stau_2}$ the decay widths can 
receive large corrections, especially for large $\tan\b$. This 
makes the perturbation expansion unreliable. In some cases the 
width can even become negative in the on--shell renormalization 
scheme. We present here a detailed description how this problem 
can be solved by an appropriate definition of the tree--level 
coupling in terms of running fermion masses and running trilinear 
coupling $A_f$ ($f = b, \tau$). For consistency between the 
on--shell and the $\overline{\rm DR}$ running parameters an 
iteration procedure is necessary. Moreover, in addition to the 
numerical results shown in \cite{0305250} we present here new 
results for $A^0 \rightarrow \st_1 \bar{\,\st_2}$, $A^0 
\rightarrow \sb_1 \!\bar{\ \sb_2}$, $A^0 \rightarrow \stau_1 
\bar{\,\stau_2}$ and the corresponding crossed channels. 
\\[3mm]
\noindent The paper is organized as follows. In section 
\ref{treelevel} we summarize the tree--level formulae. In section 
\ref{FEcorr} the full electroweak corrections are presented for 
which the explicit analytic formulae are given in the Appendices 
\ref{appVertex}, \ref{appWF}, \ref{appSE}. In section \ref{IMP} a 
detailed description of the method to improve the one--loop 
calculation is presented. In section \ref{Method} the numerical 
treatment as well as numerical results are shown. Section 
\ref{conclusions} summarizes our conclusions.

\vspace{2mm}
\section{Tree--level result}\label{treelevel}
The sfermion mixing is described by the sfermion mass matrix in
the left--right basis $(\sfL, \sfR)$, and in the mass basis
$(\sf_1, \sf_2)$, $\sf = \st, \sb$ or $\stau$,
\begin{eqnarray}
  {\cal M}_{\sf}^{\,2} \,=\,
   \left(
     \begin{array}{cc}
       m_{\sf_L}^{\,2} & a_f\, m_f
       \\[2mm]
       a_f\,m_f & m_{\sf_R}^{\,2}
     \end{array}
   \right)
  = \left( R^\sf \right)^\dag
   \left(
     \begin{array}{cc}
       m_{\sf_1}^{\,2} & 0
       \\[2mm]
       0 & m_{\sf_2}^{\,2}
     \end{array}
   \right) R^\sf \,,
\end{eqnarray}
where $R^\sf_{i\a}$ is a 2 x 2 rotation matrix with rotation angle
$\theta_{\!\sf}$, which relates the mass eigenstates $\sf_i$, $i =
1, 2$, $(m_{\sf_1} < m_{\sf_2})$ to the gauge eigenstates
$\sf_\a$, $\a = L, R$, by $\sf_i = R^\sf_{i\a} \sf_\a$ and
\begin{eqnarray}
  m_{\sf_L}^{\,2} &=& M_{\{\ti Q\!,\,\ti L \}}^2
       + (I^{3L}_f \!-\! e_{f}^{}\sin^2\!\tw)\cos2\b\,
       m_{\scriptscriptstyle Z}^{\,2}
       + m_{f}^2\,, \\[2mm]\label{MsD}
  m_{\sf_R}^{\,2} &=& M_{\{\ti U\!,\,\ti D\!,\,\ti E \}}^2
       + e_{f}\sin^2\!\tw \cos2\b\,m_{\scriptscriptstyle Z}^{\,2}
       + m_f^2\,, \\[2mm]
  a_f &=& A_f - \mu \,(\tan\b)^{-2 I^{3L}_f} \,.
\end{eqnarray}
$M_{\ti Q}$, $M_{\ti L}$, $M_{\ti U}$, $M_{\ti D}$ and $M_{\ti E}$
are soft SUSY breaking masses, $A_f$ is the trilinear scalar
coupling parameter, $\mu$ the higgsino mass parameter, $\tan\b =
\frac{v_2}{v_1}$ is the ratio of the vacuum expectation values of
the two neutral Higgs doublet states \cite{GunionHaber2,
GunionHaber1}, $I^{3L}_f$ denotes the third component of the weak
isospin of the fermion $f$, $e_f$ the electric charge in terms of
the elementary charge $e_0$, and $\tw$ is the Weinberg angle.
\\
The mass eigenvalues and the mixing angle in terms of primary
parameters are
\begin{eqnarray}
  \msf{1,2}^2
    &=& \frac{1}{2} \left(
    \msf{L}^2 + \msf{R}^2 \mp
    \sqrt{(\msf{L}^2 \!-\! \msf{R}^2)^2 + 4 a_f^2 m_f^2}\,\right) \,,
\\
  \cos\t_{\sf}
    &=& \frac{-a_f\,m_f}
    {\sqrt{(\msf{L}^2 \!-\! \msf{1}^2)^2 + a_f^2 m_f^2}}
  \hspace{2cm} (0\leq \t_{\sf} < \pi) \,,
\end{eqnarray}
and the trilinear breaking parameter $A_f$ can be written as
\begin{eqnarray}\label{mfAf}
m_f A_f  &=& m_{LR}^2 + m_f \, \mu \,(\tan\b)^{-2 I^{3L}_f}
\end{eqnarray}
with $m_{LR}^2 \equiv (m_{\sf_1}^2 - m_{\sf_2}^2) \sin \theta_\sf
\cos
\theta_\sf$. \\%
At tree--level the decay width of $A^0 \rightarrow
\tilde{f}_1 \ {\bar{\!\!\tilde{f}}}_{\!2}$ is given by
\begin{eqnarray}
\G^{\rm tree}(A^0 \rightarrow \tilde{f}_1 \
{\bar{\!\!\tilde{f}}}_{\!2}) &=& \frac{N_C^f\, \kappa (m_{A^0}^2,
m^2_{\sf_1}, m^2_{\sf_2})}{16 \,\pi\, m^3_{A^0}}\
|G_{123}^{\sf}|^2
\end{eqnarray}
with $\kappa (x, y, z) = \sqrt{(x-y-z)^2 - 4 y z}$ and the colour
factor $N_C^f = 3$ for squarks and \mbox{$N_C^f = 1$} for
sleptons, respectively. $G_{ij3}^{\sf}$ denotes the
$A^0$--$\sf_i^\ast$--$\sf_j$ coupling as given in Appendix
\ref{appCouplings}.

\vspace{2mm}
\section{Full Electroweak Corrections}\label{FEcorr}

The full one--loop corrected decay width is given by 
\begin{eqnarray}\label{1loopwidth}
\G(A^0 \rightarrow \tilde{f}_1 \ {\bar{\!\!\tilde{f}}}_{\!2}) &=&
\frac{N_C^f\, \kappa (m_{A^0}^2, m^2_{\sf_1}, m^2_{\sf_2})}{16
\,\pi\, m^3_{A^0}}\left[ |G_{123}^{\sf}|^2 + 2 {\rm Re} \left(
G_{123}^{\sf}\cdot \D G_{123}^{\sf} \right) \right] \,.
\end{eqnarray}
The (UV finite) corrections $\D G_{123}^{\sf}$ consist of the
vertex corrections $\d G_{123}^{\sf (v)}$
(Fig.~\ref{vertex-graphs}), wave--function corrections and the 
coupling counter term corrections $\d G_{123}^{\sf (c)}$ owing to 
the shift from the bare to the on--shell values, 
\begin{eqnarray}
\D G_{123}^{\sf} &=& \d G_{123}^{\sf (v)} ~+~ \d G_{123}^{\sf (w)}
~+~ \d G_{123}^{\sf (c)} \,.
\end{eqnarray}
The  renormalization procedure with the fixings of the counter
terms is given in \cite{0305250}. The explicit formulae of the 
vertex corrections $\d G_{123}^{\sf (v)}$ as well as the various 
contributions to the wave--function corrections $\d G_{123}^{\sf 
(w)}$, 
\begin{eqnarray}
\delta G_{123}^{\sf (w)} &=& \frac{1}{2}\Re\Big[\d Z_{11}^{\sf} + 
\d Z_{22}^{\sf} + \d Z_{33}^{H} \Big] G_{123}^\sf \,, 
\end{eqnarray}
can be found in the Appendices \ref{appVertex} and \ref{appWF}. 
For the explicit formulae of the self--energies needed for the 
calculation of the counter term correction, $\d G_{123}^{\sf (c)}$ 
(see eq.~(23) of \cite{0305250}) , we refer to the Appendix 
\ref{appSE}. 
\\ Due to the diagrams with photon exchange we also have to
consider real photon emission corrections to cancel the infrared
divergences (Fig.~\ref{vertex-graphs}). Therefore the corrected
(UV-- and IR--convergent) decay width is
\begin{eqnarray}\label{correctedwidth}
\G^{\rm corr}(A^0 \rightarrow \tilde{f}_1 \
{\bar{\!\!\tilde{f}}}_{\!2}) &\equiv& \G(A^0 \rightarrow
\tilde{f}_1 \ {\bar{\!\!\tilde{f}}}_{\!2}) \,+\, \G(A^0
\rightarrow \tilde{f}_1 \ {\bar{\!\!\tilde{f}}}_{\!2}\,\g)\,.
\end{eqnarray}
Throughout the paper we use the SUSY invariant dimensional
reduction $(\overline{\rm DR})$ as regularization scheme. For
convenience we perform the calculation in the 't Hooft--Feynman
gauge, $\xi=1$.

\section{Improvement of One--loop Corrections}\label{IMP}
It has already pointed out in \cite{0305250} that in the case of 
bottom squarks and tau sleptons, especially for large $\tan\b$ the 
corrections to the decay widths $A^0 \rightarrow \sb_1 \!\bar{\ 
\sb_2}$ and $A^0 \rightarrow \stau_1 \bar{\,\stau_2}$ can be very 
large in the on--shell renormalization scheme. If the corrections 
are negative, the one--loop corrected width can even become 
negative and therefore unphysical. Hence the perturbation 
expansion around the on--shell tree--level is no longer reliable. 
It has been shown in \cite{dmb, impSUSYQCD} that, in the case of 
the decays into bottom squarks, the source of these large 
corrections are mainly the counter terms for $m_b$ and the 
trilinear coupling $A_b$, in particular the SUSY--QCD corrections. 
However, despite the absence of strong interactions for the decay 
into tau sleptons, the corrections become extremely large. We show 
that this problem can be solved by defining an appropriate 
tree--level in terms of running values for $m_f$ and $A_f$. The 
expansion around this new tree--level then no longer suffers from 
bad convergence.\\

\noindent{\bf Correction to {\boldmath{$m_b$}\label{sec:mbMSSM}}}
\\ %
First we review the improvement of the perturbation expansion by
using $\overline{\rm DR}$ running bottom quark masses, following
\cite{impSUSYQCD, BL, hdecay}.

If the Yukawa coupling $h_b$ is given at tree--level in terms of
the pole mass $m_b$, the one--loop corrections become very large
due to gluon and gluino exchange contributions to the counter term
$\d m_b$. The large counter term caused by the gluon loop is
absorbed by using SM 2--loop renormalization group equations in
the $\overline{\rm MS}$ scheme \cite{impSUSYQCD, BL, hdecay}. Thus
we obtain the SM running bottom $\hat m_b(Q)_{\rm SM}$:
\begin{equation}
\hat m_b(Q)_{\rm SM}^{\overline{\rm MS}}=\left(\frac{\hat
m_b(Q)_{\rm SM}^{\overline{\rm MS}}}{\hat m_b(m_b)_{\rm
SM}^{\overline{\rm MS}}}\right) \, \hat m_b(m_b)_{\rm
SM}^{\overline{\rm MS}}
\end{equation}
The ratio $\left(\hat m_b(Q)_{\rm SM}^{\overline{\rm MS}}/\hat
m_b(m_b)_{\rm SM}^{\overline{\rm MS}}\right)$ can be expressed as
\begin{eqnarray*}
\frac{\hat m_b(Q)_{\rm SM}^{\overline{\rm MS}}}{\hat m_b(m_b)_{\rm
SM}^{\overline{\rm MS}}} &=& \left\{
  \begin{array}{l@{\qquad\qquad}l}
    {\displaystyle{
    \frac{c_5(\alpha_s^{(2)}(Q)/\pi)}{c_5(\alpha_s^{(2)}(m_b)/\pi)}
    }}
    & (m_b < Q \le m_t) \,,
    \\[7mm]
    {\displaystyle{
    \frac{c_6(\alpha_s^{(2)}(Q)/\pi)}{c_6(\alpha_s^{(2)}(m_t)/\pi)}\,
    \frac{c_5(\alpha_s^{(2)}(m_t)/\pi)}{c_5(\alpha_s^{(2)}(m_b)/\pi)}}}
    & (Q > m_t) \,,
  \end{array}
\right.
\end{eqnarray*}
where we have used the functions
\begin{eqnarray*}
  \begin{array}{l@{\qquad\qquad}l}
    {\displaystyle{
    c_5(x) ~=~ \left(\frac{23}{6}x\right)^{\frac{12}{23}}(1+1.175x)}}
    & (m_b < Q \le m_t)\,,
    \\[5mm]
    {\displaystyle{
    c_6(x) ~=~ \left(\frac{7}{2}x\right)^{\frac{4}{7}}(1+1.398x)}}
    & (Q > m_t)\,,
  \end{array}
\end{eqnarray*}
and the 2--loop RGEs for $\a_s$ \cite{hdecay}, 
\begin{eqnarray}
\alpha_s^{(2)}(Q) = \frac{12\pi}{(33-2\,n_f)\ln \frac{Q^2}
{\Lambda^2_{n_f}}}
\left( 1-\frac{6(153-19\,n_f)}{(33-2\,n_f)^2}
\frac{\ln\ln\frac{Q^2}{\Lambda^2_{n_f}}}{\ln\frac{Q^2}{\Lambda^2_{n_f}}}
\right)\, ,
\end{eqnarray}
with $n_f=5$ or 6 for $m_b < Q \le m_t$ or $Q > m_t$,
respectively. For the SM $\overline{\rm DR}$ running bottom mass
at the scale $Q=m_b$ we use the $\overline{\rm MS}$ equation
\begin{equation}
\hat m_b(m_b)_{\rm SM}^{\rm \overline{\rm MS}} = m_b
\left[1+\frac{4}{3}\frac{\alpha_s^{(2)}(m_b)}{\pi}
+K_q\left(\frac{\alpha_s^{(2)}(m_b)}{\pi}\right)^2\right]^{-1}\, ,
\end{equation}
with $K_q=12.4$ and then convert to $\overline{\rm DR}$ using
one--loop running $\a_s (Q)$:
\begin{eqnarray}
\hat m_b(Q)_{\rm SM} &=& \left(\frac{\hat m_b(Q)_{\rm
SM}^{\overline{\rm MS}}}{\hat m_b(m_b)_{\rm SM}^{\overline{\rm
MS}}}\right) \, \hat m_b(m_b)_{\rm SM}^{\rm \overline{\rm MS}} ~-~
\frac{\a_s(Q)}{3\pi}\,m_b
\end{eqnarray}
In the MSSM, for large $\tan\b$ the counter term to $m_b$ can be
very large due to the gluino--mediated graph \cite{dmb, hbbnew,
chankowski}. Here we absorb the gluino contribution as well as the
sizeable contributions from neutralino and chargino loops and the
remaining electroweak self--energies into the
Higgs--sfermion--sfermion tree--level coupling. In such a way we
obtain the full $\overline{\rm DR}$ running bottom quark mass
\begin{eqnarray}\label{mbMSSM}
   \hat m_b(Q)_{\rm MSSM} &=& \hat m_b(Q)_{\rm SM} + \d m_b(Q) \,.
\end{eqnarray}
The explicit form of the electroweak contribution to the counter
term $\d m_b(Q)$ is given in Appendix~\ref{appfermion-SE}. \\

\noindent{\bf Correction to {\boldmath{$A_{b,\tau}$}}}\\ The
second source of a very large correction (in the on--shell scheme)
are the counter terms for the trilinear coupling $A_{b,\tau}$,
(see eq.~\ref{mfAf}),
\begin{eqnarray}
\d A_{b,\tau} &=& \frac{\d m_{LR}^2}{m_{b,\tau}} ~-~
\frac{m_{LR}^2}{m_{b,\tau}}\, \frac{\d m_{b,\tau}}{m_{b,\tau}} ~+~
\d\mu \tan\b ~+~ \mu\, \d\!\tan\b \,.
\end{eqnarray}
Again, the big bottom mass correction $\d m_b$ contributes to $\d
A_b$, but also the counter term of the left--right mixing elements
of the sfermion mass matrix, $\d m_{LR}^2$, gives a very large
correction for higher values of $\tan\b$. In particular, in the
case of the decay into staus, this is the main source for the bad
convergence of the tree--level expansion. As in the case of the
large correction to $m_b$ we redefine the
Higgs--sfermion--sfermion tree--level coupling in terms of
$\overline{\rm DR}$ running $\hat A_{b,\tau} (m_{A^0})$ . Because
of the fact that the counter terms $\d A_{b,\tau}$ (for large
$\tan\b$) can become several orders of magnitude larger than the
on--shell $A_{b,\tau}$ we use $\hat A_{b,\tau} (m_{A^0})$ as input
\cite{impSUSYQCD}. In order to be consistent we have to perform an
iteration procedure to get all the correct running and on--shell
masses, mixing angles and other parameters. This procedure is
described below.

\section{Method of improvement}\label{Method}
In this section we will explain in detail how we can improve the
perturbation calculation for the sbottom and stau case by using
$\overline{\rm DR}$ running values for $m_b$ and $A_{b,\tau}$ in
the Higgs--sfermion--sfermion tree--level couplings. Since we take
$\overline{\rm DR}$ running values for $\hat A_b$ and $\hat
A_\tau$ as input and all other parameters on--shell we will have
to pay attention to the sbottom and stau sector in order to get
consistently all needed running and on--shell masses, mixing
angles and other parameters. Here we adopt the procedure developed
in \cite{impSUSYQCD} and also extend it to the electroweak case.

\subsection{Calculation of running and on--shell parameters \label{iteration}}

{\bf Stop sector:}\\ We start our calculation in the stop sector.
Because all input parameters in the stop sector are on--shell we
obtain the on--shell masses $m_{\st_1}$, $m_{\st_2}$ and the stop
mixing angle $\theta_\st$ by diagonalizing the stop mass matrix in
the $\st_L$--$\st_R$ basis, see chapter~\ref{treelevel}. The
running stop masses $\hat m_{\st_i}$ and mixing angle $\hat
\theta_\st$ are calculated at the scale $Q = Q_\st =
\sqrt{m_{\st_1}m_{\st_2}}$ by adding the appropriate counter terms
to the on--shell values in
\begin{eqnarray}
  \hat m_{\st_i}^2 (Q_\st) &=& m_{\st_i}^2 + \d m_{\st_i}^2\,,
  \\
  \d m_{\st_i}^2 &=& \Re\, \Pi_{ii}^\st (m_{\st_i}^2) \,,
  \\[2mm]
  \hat \theta_\st (Q_\st) &=& \theta_\st + \d\theta_\st\,.
\end{eqnarray}
The electroweak parts of the sfermion self--energies $\Pi_{ii}^\sf
(m_{\sf_i}^2)$ are given in Appendix~\ref{appsfermion-SE} and the
SUSY--QCD contributions, $\Pi_{ii}^{\rm SUSY-QCD}(m_{\sf_i}^2)$
are given in eqs.~(25)--(27) in \cite{SUSY-QCD}. Here and in the
following all running parameters $\hat X (Q)$ are related to their
on--shell values $X$ by $\hat X (Q) = X + \d X$, with $\d X$ being
the full one--loop counter term --- also including the SUSY--QCD
parts.  The mixing angle is fixed by \cite{sqyukawa}
\begin{eqnarray}
   \delta \theta_{\sf} & = & \frac{1}{4}\, \left(
   \d Z^\sf_{12} - \d Z^\sf_{21}\right)
   = \frac{1}{2\big(m_{\sf_1}^2 \!-\! m_{\sf_{2}}^2\big)}\, {\rm Re}
   \left( \Pi_{12}^\sf(m_{\sf_{2}}^2) + \Pi_{21}^\sf(m_{\sf_{1}}^2)
    \right) \,.
\end{eqnarray}
For $\overline{\rm DR}$ running $\hat m_t$ we use the formulae 
from section~\ref{sec:mbMSSM} with the obvious substitutions $m_b 
\rightarrow m_t$ and $K_q = 10.9$ for the top--case. Next we 
evaluate the running parameters $\hat M_{\ti Q}(Q)$ and $\hat 
M_{\ti U}(Q)$ by inserting the running values $\hat m_{\st_i}^2 
(Q), \hat \theta_\st (Q), \hat m_t(Q)_{\rm MSSM}$, $\hat 
m_{\scriptscriptstyle Z}(Q) = m_{\scriptscriptstyle Z} + \d 
m_{\scriptscriptstyle Z}$, $\hat\b(Q) = \b + \d\b$ and 
$\hat{\theta}_{\scriptscriptstyle W} = \tw 
-\frac{1}{\sin\tw}\left( \frac{\d m_{\scriptscriptstyle 
W}}{m_{\scriptscriptstyle W}} -\frac{\d m_{\scriptscriptstyle 
Z}}{m_{\scriptscriptstyle Z}} \right)$ into the equations 
\begin{eqnarray}\label{msQ2}
M_{\ti Q}^{2} & = & m_{\st_1}^2 \cos^{2}\theta_{\st} + m_{\st_2}^2
\sin^{2}\theta_{\st} - m_{t}^{2} - m_{\scriptscriptstyle Z}^2\cos
2\beta \left(I^{3L}_t \!-\!e_t\sin^2\tw\right)\,,
\\[2mm]
M_{\ti U}^{2} & = & m_{\st_1}^2 \sin^{2}\theta_{\st} + m_{\st_2}^2
\cos^{2}\theta_{\st} - m_{t}^{2} - m_{\scriptscriptstyle Z}^2\cos
2\beta \,e_t\sin^2 \tw \, .
\end{eqnarray}
For the running value of $A_t$ we use (see eq.~(\ref{mfAf}))
\begin{eqnarray}\label{Atr}
  \hat A_t &=& (\hat m_{\st_1}^2 - \hat m_{\st_2}^2)\,
  \frac{\sin 2\hat \theta_\st}{\hat m_t}
  + \hat\mu \,\cot\hat\beta \,,
\end{eqnarray}
where we have taken running $\hat \mu(Q) = \mu + (\d X)_{22}$
\cite{0104109, Willi}. \\ 

\noindent{\bf Sbottom sector:}\\ In the sbottom sector we have
given all parameters on--shell except the parameter for the
trilinear coupling, $\hat A_b (Q)$, which is running. First we
calculate $\hat m_b(Q_\sb)_{\rm MSSM}$ from eq.~(\ref{mbMSSM}) at
the scale $Q_\sb = \sqrt{m_{\sb_1}m_{\sb_2}}$ . From the stop
sector we already know the running values of $M_{\ti Q}, \tan\b$
and $\mu$. Then we diagonalize the sbottom mass matrix using
${\hat m}_b(Q_\sb)_{\rm MSSM}, {\hat M}_{\ti Q}, \tan\hat\b,
\hat\mu$ and on--shell $M_{\ti D}$, which is near its running
value $\hat M_{\ti D}$, to obtain the starting values for ${\hat
m}_{\sb_i}$ and ${\hat\theta}_\sb$. The on--shell sbottom masses
$m_{\sb_i}$ and the mixing angle $\theta_\sb$ are calculated from
their running values by subtracting the appropriate counter terms,
i.~e. $m_{\sb_i}^2 = \hat m_{\sb_i}^2(Q) - \d m_{\sb_i}^2$,
$\theta_\sb = \hat \theta_\sb (Q) - \d \theta_\sb$. Now we can
compute the running value for $M_{\ti D}$. Using the relation (see
eq.~(\ref{MsD}))
\begin{eqnarray}
M_{\ti D}^2 &=& m_{\sb_1}^2 \sin^2\theta_\sb + m_{\sb_2}^2
\cos^2\theta_\sb - m_b^2 - m_{\scriptscriptstyle Z}^2 \cos 2\b\,
e_b \sin^2\tw
\end{eqnarray}
we get $\hat M_{\ti D} = (M_{\ti D}^2 + \d M_{\ti D}^2)^{1/2}
\approx M_{\ti D} + \d M_{\ti D}^2/(2 M_{\ti D})$ with
\begin{eqnarray}\non
\d M_{\ti D}^2 &=& \d m_{\sb_{1}}^2 \sin^2\theta_\sb + \d
m_{\sb_{2}}^2 \cos^2\theta_\sb + \Big( m_{\sb_1}^2 \!-\!
m_{\sb_2}^2 \Big) \sin 2\theta_\sb \, \d\theta_\sb - 2 m_b \,\d
m_b
\\ \non
&& - \d m_{\scriptscriptstyle Z}^2 \cos 2\b\, e_b \sin^2\tw + 2
m_{\scriptscriptstyle Z}^2 \sin 2\b \, \d\b\, e_b \sin^2\tw -
m_{\scriptscriptstyle Z}^2 \cos 2\b\, e_b \,\d\!\sin^2 \tw
\\
\end{eqnarray}
and $\d m_b = \hat m_b(Q_\sb)_{\rm MSSM}\!-\!m_b$. Because the
parameters involved in these calculations are very entangled,
e.~g. $\hat M_{\ti D}$ depends on the $\d m_{\sb_i}$ which
themselves depend on $M_{\ti D}$, we have to perform an iteration
procedure. \\

\begin{figure}[h!]
\begin{picture}(160,150)(0,0)
    \put(10,5){\mbox{\resizebox{14.1cm}{!}
    {\includegraphics{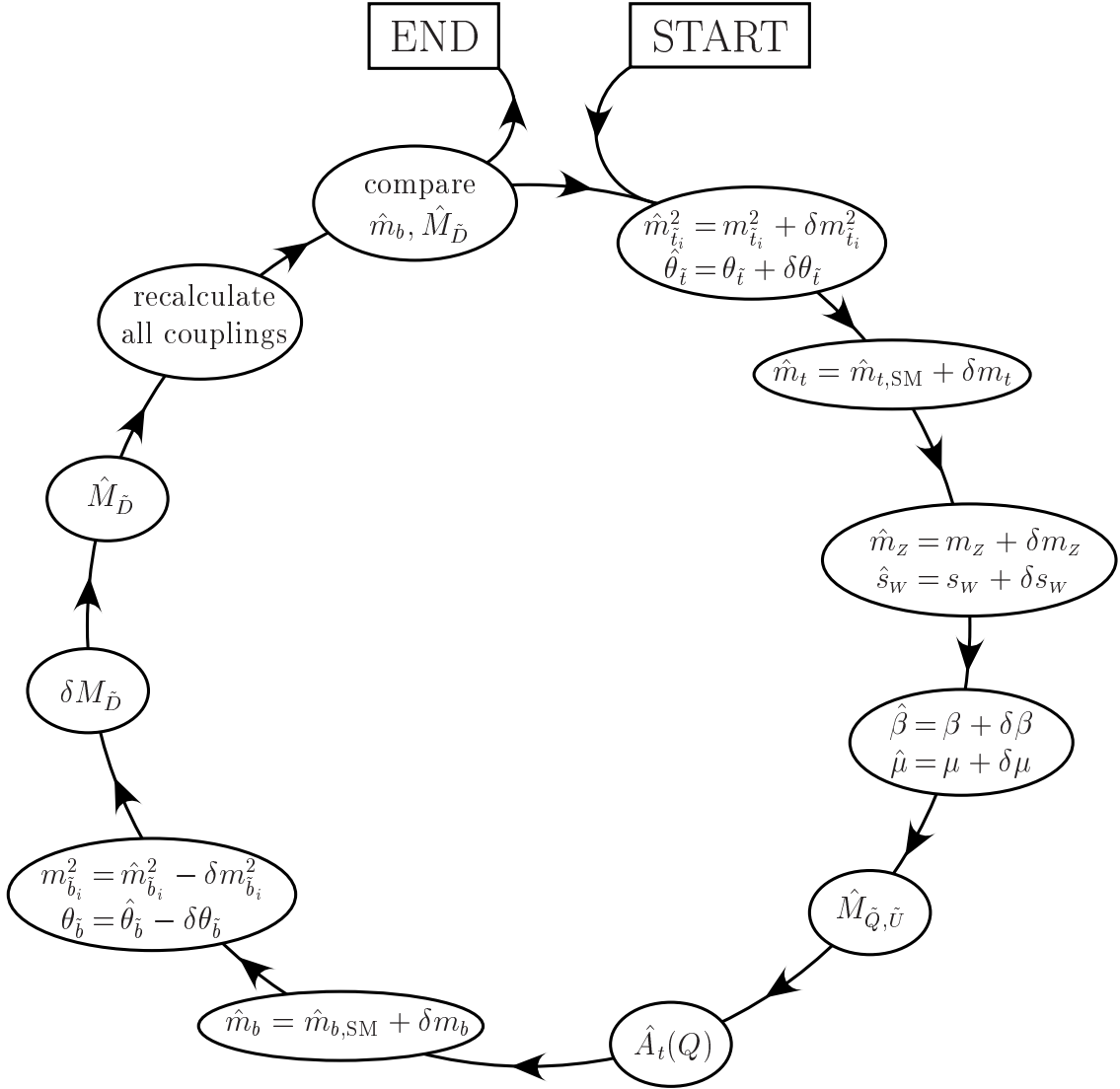}}}}
\end{picture}
\caption{Simplified flowchart for the iteration procedure. For
details see section~\ref{iteration}. \label{flowchart}}
\end{figure}

\noindent{\bf Iteration procedure:}\\ Here we will describe in
detail the procedure how we obtain all necessary on--shell and
running parameters. For convenience, we shortly denote all masses,
parameters, couplings etc. for a certain $n \geq 1$ in the
iteration by ${\hat{\mathcal X}}^{(n)}$. As starting values
${\hat{\mathcal X}}^{(0)}$ we take on--shell masses and parameters
(except $\hat A_b$ which is running) and the couplings derived
from these quantities. The only exceptions are the standard model
running fermion masses $\hat m_f^{(0)} = \hat m_f (Q)_{\rm SM}$.
$\hat m_f$ shortly stands for the full $\overline{\rm DR}$ running
fermion masses, $\hat m_f (Q)_{\rm MSSM}$.
\\ The single steps of the iteration procedure are the following:
\begin{enumerate}
\item
The running stop masses and the stop mixing angle are calculated
as explained\\[2mm] above by $\hat m_{\st_i}^{2\, (n)} =
m_{\st_i}^{2} + \d m_{\st_i}^{2\, (n)} \big( {\hat{\mathcal
X}}^{(n-1)} \big)$ and $\hat \theta_\st^{(n)} = \theta_\st + \d
\theta_\st^{(n)} \big( {\hat{\mathcal X}}^{(n-1)} \big)$.
\item
$\hat m_t^{(n)} = \hat m_{t, \rm SM} + \d m_t^{(n)}\big(
{\hat{\mathcal X}}^{(n-1)} \big)$
\item
$\hat m_{\scriptscriptstyle Z}^{(n)} = m_{\scriptscriptstyle Z} +
\d m_{\scriptscriptstyle Z}^{(n)}\big( {\hat{\mathcal X}}^{(n-1)}
\big)$ and
\\[2mm]
$\sin^2\hat\tw^{(n)} = \sin^2\tw + \d\!\sin^2\tw^{(n)}$ with
$\d\!\sin^2\tw^{(n)} = -\cos^2\tw \left( {\displaystyle{\frac{\d
m_{\scriptscriptstyle W}}{m_{\scriptscriptstyle W}} - \frac{\d
m_{\scriptscriptstyle Z}}{m_{\scriptscriptstyle Z}}}} \right)\big(
{\hat{\mathcal X}}^{(n-1)} \big)$
\item
The running value of $\tan\b$, $\tan\hat\b^{(n)} = \tan\b + 
\d\tan\b^{(n)}$,\\[2mm] with $\d\tan\b^{(n)} = 
{\displaystyle{\frac{1}{m_{\scriptscriptstyle Z} \sin 2\b}}}\, 
{\rm Im} \Pi_{A^0 Z^0} \big( {\hat{\mathcal X}}^{(n-1)} \big) 
\tan\b$ \cite{pokorski}. 
\item
$\hat \mu^{(n)} = \mu + \d\mu^{(n)}$ with $\d\mu^{(n)} = \d
X_{22}\big({\hat{\mathcal X}}^{(n-1)}\big)$.
\item
The soft SUSY breaking masses $\hat M_{\ti Q, \ti U}^{(n)}$ are
calculated from $\hat m_{\st_i}^{(n)}$, $\hat \theta_\st^{(n)}$,
${\hat m}_t(Q_{\st})^{(n)}$, $\hat m_{\scriptscriptstyle
Z}^{(n)}$, $\sin^2\hat\tw^{(n)}$ and $\tan\hat\b^{(n)}$.
\item
We compute the running $\hat A_t$ by using running values in
eq.~(\ref{Atr}):\\[2mm] $\hat A_t^{(n)} = \left(\hat
m_{\st_1}^{2\, (n)} - m_{\st_2}^{2\, (n)}\right)\,
{\displaystyle{\frac{\sin 2\hat\theta_\st^{(n)}}{\hat m
_t^{(n)}}}}
+ \hat\mu^{(n)} \,\cot\hat\beta^{(n)}$
\item\label{sbstepstart}
In the sbottom sector we obtain $\d m_b^{(n)}$ from the running
values already calculated in steps 1.--7., like $\hat
m_{\sb_i}^{(n)}$, $\hat \theta_\sb^{(n)}$ or $\hat m_f^{(n)}$, and
the remaining masses, couplings etc. from ${\hat{\mathcal
X}}^{(n-1)}$.
\item
$\hat m_b^{(n)} = \hat m_{b,\rm SM} + \d m_b^{(n)}$\,.
\item
We receive the running sbottom masses, $\hat m_{\sb_i}^{(n)}$, and
the mixing angle, $\hat \theta_\sb^{(n)}$, by solving the mass
eigenvalue problem with the running values of $\hat M_{\ti
Q}^{(n)}$, $\hat M_{\ti D}^{(n-1)}$, $\hat m_b^{(n)}$, $\hat A_b$,
$\hat \mu^{(n)}$ and $\tan{\hat\b}^{(n)}$.
\item
The on--shell sbottom masses $m_{\sb_i}^{2\, (n)} = {\hat
m}_{\sb_i}^{2\, (n)} - \d m_{\sb_i}^{2\, (n)} (Q_\sb^{(n)})$ at
the scale $Q_\sb^{(n)} = \sqrt{\hat m_{\sb_1}^{(n)} \hat
m_{\sb_2}^{(n)}}$, and $\theta_\sb^{(n)} = \hat \theta_\sb^{(n)} -
\d\theta_\sb^{(n)}$\,.
\item
$ \d M_{\ti D}^{2\, (n)} ~=~ \d m_{\sb_{1}}^{2\, (n)}
\sin^2\theta_\sb + \d m_{\sb_{2}}^{2\, (n)} \cos^2\theta_\sb +
\Big( m_{\sb_1}^2 \!-\! m_{\sb_2}^2 \Big) \sin 2\theta_\sb \,
\d\theta_\sb^{(n)} - 2 m_b \,\Big( {\hat m}_b^{(n)}-m_b \Big)$
\\[2mm] \hphantom{aaaaaaa} $- \d m_{\scriptscriptstyle Z}^{2\, (n)}
\cos 2\b\, e_b \sin^2\tw + 2 m_{\scriptscriptstyle Z}^2 \sin 2\b
\, \d\b^{(n)}\, e_b \sin^2\tw - m_{\scriptscriptstyle Z}^2 \cos
2\b\, e_b \,\d\!\sin^2 \tw^{(n)}$.\\[2mm] (Remember that the
values without a hat ($\hat{\hphantom{a}}$) are on--shell ones!)
\item\label{sbstepend}
${\displaystyle{{\hat M}_{\ti D}^{(n)} = M_{\ti D} + \frac{1}{2}
\frac{\d M_{\ti D}^{2\, (n)}}{M_{\ti D}}}}$\,.
\item
In the sneutrino sector we calculate the running sneutrino mass
\\[2mm] $\hat m_{\ti\nu_\tau}^{2\, (n)} = m_{\ti\nu_\tau}^{2} + \d
m_{\ti\nu_\tau}^{2\, (n)} \big( {\hat{\mathcal X}}^{(n-1)} \big)$
and $\hat M_{\ti L}^{2\, (n)} = \hat m_{\ti\nu_\tau}^{2\, (n)} -
\frac{1}{2}\, \hat m_{\scriptscriptstyle Z}^{(n)} \cos 2
\hat\beta^{(n)}$, see also eq~(\ref{msQ2}).
\item
In the stau sector the values for running $\hat m_\tau^{(n)}, \hat
m_{{\ti\tau}_i}^{(n)}$ etc. are calculated like in the steps
\ref{sbstepstart}--\ref{sbstepend} in the sbottom sector with the
evident substitution $\sb \rightarrow \stau$ for the corresponding
parameters and $M_{\ti Q} \rightarrow M_{\ti L}, M_{\ti D}
\rightarrow M_{\ti E}$.
\item
All couplings are recalculated with the new running parameters
$\rightarrow {\hat{\mathcal X}}^{n}$.
\end{enumerate}
The iteration starts with $n=1$ and ends, when certain parameters
are calculated precisely enough for a given accuracy, i.~e.
$\left| 1 - \frac{\hat x^{(n)}}{\hat x^{(n-1)}} \right| <
\varepsilon$ for $\hat x = \{ \hat m_b, \hat M_{\ti D}, \hat
m_\tau, \hat M_{\ti E} \}$. For $\varepsilon$ we choose
$\varepsilon = 10^{-8}$. We have checked the consistency of this
procedure by computing the on--shell $M_{\ti D}$ and running
$M_{\ti Q}$ from the sbottom sector by using
\begin{eqnarray}
M_{\ti D}^{2} &=&  m_{\sb_1}^2 \sin^{2}\theta_{\sb} + m_{\sb_2}^2
\cos^{2}\theta_{\sb} - m_{b}^{2} - m_{\scriptscriptstyle Z}^2\cos
2\beta \,e_b\sin^2 \tw \,,
\\
\hat M_{\ti Q}^{2} & = & \hat m_{\sb_1}^2 \cos^{2}\hat
\theta_{\sb} + \hat m_{\sb_2}^2 \sin^{2}\hat \theta_{\sb} - \hat
m_{b}^{2} - \hat m_{\scriptscriptstyle Z}^2\cos 2\hat \beta
\left(I^{3L}_b \!-\!e_b\sin^2\hat\tw\right)\,,
\end{eqnarray}
which are equal (up to higher order corrections) to the on--shell
input $M_{\ti D}$ and running $M_{\ti Q}$ from the stop sector. \\
For easier reading the single steps of the iteration procedure of
the stop and sbottom sector are depicted in the flowchart in
Fig.~\ref{flowchart}.
\clearpage

\subsection{Numerical results}
In the following numerical examples, we take for the standard 
model parameters $m_{\scriptscriptstyle Z} = 91.1876$~GeV, 
$m_{\scriptscriptstyle W} = 80.423$~GeV, $\sin^2 \tw = 1 - 
m_{\scriptscriptstyle W}^2/m_{\scriptscriptstyle Z}^2$, $\a 
(m_{\scriptscriptstyle Z}) = 1/127.934$, $m_t = 174.3$~GeV, $m_b = 
4.7$~GeV, $m_\tau = 1.8$~GeV and $\{m_u, m_d, m_e, m_c, m_s, 
m_\mu\} = \{4, 8, 0.511,$ $1300, 200, 106\}$~MeV for $1^{\rm st}$ 
and $2^{\rm nd}$ generation fermions. $M'$ is fixed by the gaugino 
unification relation $M' = {\displaystyle{\frac{5}{3}}} \, 
\tan^2\tw M$, therefore the gluino mass is related to $M$ by 
$m_{\tilde g} = (\a_s(m_{\tilde g})/\a)\sin^2\tw M$. In order to 
reduce the number of parameters in the input parameter set, we 
assume $M_{{\ti Q}} \equiv M_{{\ti Q}_{3}} = \frac{10}{9} M_{{\ti 
U}_{3}} = \frac{10}{11} M_{{\ti D}_{3}} = M_{{\ti L}_{3}} = 
M_{{\ti E}_{3}} = M_{{\ti Q}_{1,2}} = M_{{\ti U}_{1,2}} = M_{{\ti 
D}_{1,2}} = M_{{\ti L}_{1,2}} = M_{{\ti E}_{1,2}}$ for the first, 
second and third generation soft SUSY breaking masses as well as 
$A \equiv A_t = A_b = A_\tau$ for all (s)fermion generations, if 
not stated otherwise. \\

\noindent {\bf stop--case:}

\begin{figure}[h!]
\begin{picture}(160,65)(0,0)
    \put(30,5){\mbox{\resizebox{8.6cm}{!}
    {\includegraphics{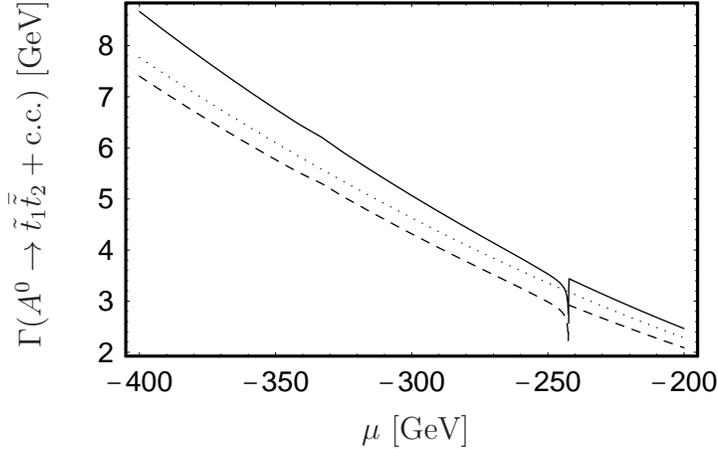}}}}
    \put(74,3){\makebox(0,0)[t]{{$\mu$ [GeV]}}}
    \put(20,12){\rotatebox{90}{{{$\Gamma (A^0 \rightarrow
                        \st_1 \!\bar{\,\st_2}+{\rm c.c.})$\ [GeV]}}}}
\end{picture}\caption{Tree--level (dotted line), full electroweak
corrected decay width (dashed line) and full one--loop
(electroweak and SUSY--QCD) corrected width (solid line) of $A^0
\rightarrow \st_1 \!\bar{\,\st_2}$ as a function of $\mu$.
\label{mudep_st_os}}
\end{figure}

In Fig.~\ref{mudep_st_os} we show the tree--level and the
corrected widths to $A^0 \rightarrow \st_1 \!\bar{\,\st_2}$ for
$\tan\b = 15$ and $\{ m_{A^0}, A, M, M_{\ti Q} \} = \{700, -500,
120, 300\}~\mbox{GeV}$ as a function of the higgsino mass
parameter $\mu$. The electroweak corrections are almost constant
about $-7$\%. \\ At $\mu \approx -242~\mbox{GeV}$ one can identify
the pseudo--threshold coming from $\st_2 \rightarrow t\, \nt_4$.

\begin{figure}[h!]
\begin{picture}(160,65)(0,0)
    \put(29,5){\mbox{\resizebox{9cm}{!}
    {\includegraphics{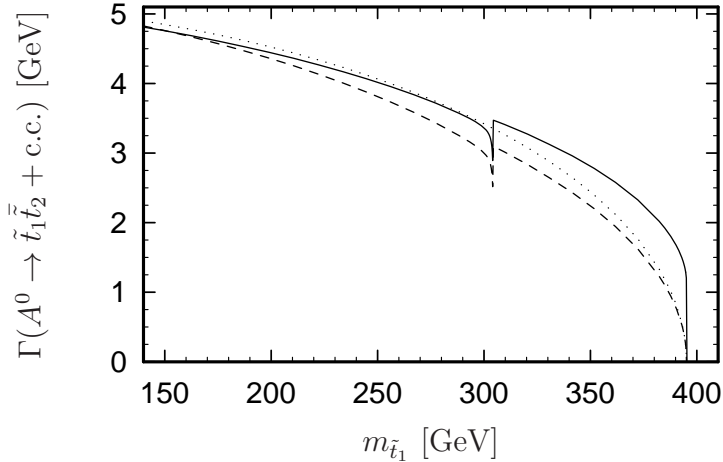}}}}
    \put(76,3){\makebox(0,0)[t]{{$m_{\st_1}$ [GeV]}}}
    \put(20,12){\rotatebox{90}{{{$\Gamma (A^0 \rightarrow
                          \st_1 \!\bar{\,\st_2}+{\rm c.c.})$ [GeV]}}}}
\end{picture}\caption{Tree--level (dotted line), full electroweak
corrected decay width (dashed line) and full one--loop
(electroweak and SUSY--QCD) corrected width (solid line) of $A^0
\rightarrow \st_1 \!\bar{\,\st_2}$ as a function of $m_{\st_1}$.
\label{mst1dep_st_os}}
\end{figure}

Fig.~\ref{mst1dep_st_os} shows the tree--level, the full
electroweak and the full one--loop corrected (electroweak and
SUSY--QCD) decay width of $A^0 \rightarrow \st_1 \!\bar{\,\st_2}$
as a function of the lighter stop mass, $m_{\st_1}$, where $M_{\ti
Q}$ is varied from 200 to 450 GeV. As input parameters we choose
$\{ m_{A^0}, \mu, A, M \} = \{900, 250, 300, 120\}~\mbox{GeV}$ and
$\tan\b = 7$. Again, in a large region of the parameter space the
electroweak corrections are comparable to the SUSY--QCD ones. The
pseudo--threshold at $m_{\st_1} \approx 304~\mbox{GeV}$ originates
from $\st_2 \rightarrow t\, \nt_3$ in the wave--function 
correction. 

\begin{figure}[h!]
\begin{picture}(160,65)(0,0)
    \put(28.5,5){\mbox{\resizebox{8.8cm}{!}
    {\includegraphics{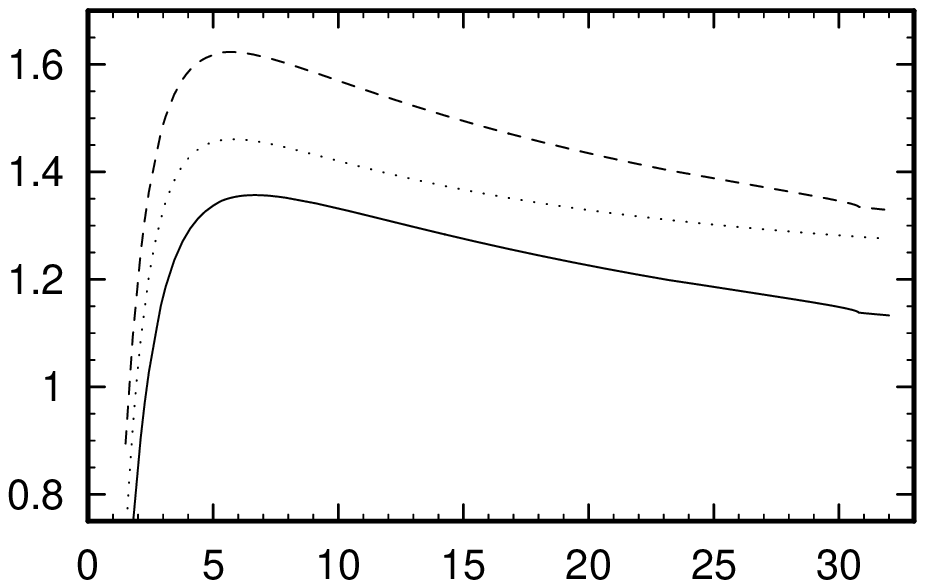}}}}
    \put(76,3){\makebox(0,0)[t]{{$\tan\b$}}}
    \put(20,20){\rotatebox{90}{{{$\Gamma (\st_2 \rightarrow
                                  \st_1 A^0)$ [GeV]}}}}
\end{picture}
\caption{$\tan\b$--dependence of the tree--level (dotted line),
full electroweak corrected (dashed line) and full one--loop
corrected (solid line) decay widths of $\st_2 \rightarrow \st_1
A^0$. \label{tanbdep_st_os_crossed}}
\end{figure}

In Fig.~\ref{tanbdep_st_os_crossed} the dependence of the crossed
channel decay width, $\Gamma (\st_2 \rightarrow \st_1 A^0)$, as a
function of $\tan\b$ is given. We see that the electroweak
corrections have different sign compared to the SUSY--QCD ones and
go up to 10\%. As input parameters we have chosen $\{ m_{A^0},
\mu, A, M, M_{\ti Q} \} = \{170, 500, -390, 250, 350\}~\mbox{GeV}$
as well as $M_{\ti U_3} = 450~\mbox{GeV}$ to get an acceptable
stop mass splitting.

\begin{figure}[h!]
\begin{picture}(160,65)(0,0)
    \put(28.5,5){\mbox{\resizebox{8.6cm}{!}
    {\includegraphics{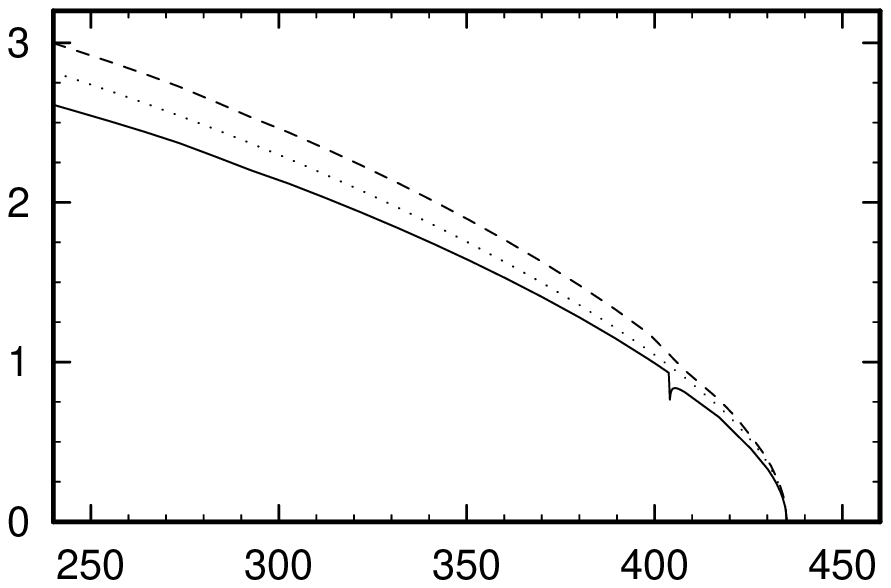}}}}
    \put(76,3){\makebox(0,0)[t]{{$m_{\st_1}$ [GeV]}}}
    \put(20,20){\rotatebox{90}{{{$\Gamma (\st_2 \rightarrow
                                  \st_1 A^0)$ [GeV]}}}}
\end{picture}
\caption{$m_{\st_1}$--dependence of the tree--level (dotted line),
full electroweak corrected (dashed line) and full one--loop
corrected (solid line) decay widths of $\st_2 \rightarrow \st_1
A^0$. \label{mst1dep_st_os_crossed}}
\end{figure}

Fig.~\ref{mst1dep_st_os_crossed} shows the decay width $\Gamma 
(\st_2 \rightarrow \st_1 A^0)$ as a function of $m_{\st_1}$, 
varying $M_{\ti Q_3}$ from 200 to 460 GeV. To get a larger mass 
splitting for the top squarks, we relax the conditions for 3rd 
generation squarks and take $\{ M_{\ti U_3}, M_{\ti D_3} \} = \{ 
500, 300 \}~\mbox{GeV}$. All other SUSY breaking masses are fixed 
at 300 GeV. For the remaining input parameters we choose $\{ 
m_{A^0}, \mu, A \} = \{ 120, -400, -350 \}~\mbox{GeV}$ and $\tan\b 
= 7$. 

\noindent {\bf sbottom--case:}

\begin{figure}[h!]
\begin{picture}(160,65)(0,0)
    \put(26,5){\mbox{\resizebox{8.8cm}{!}
    {\includegraphics{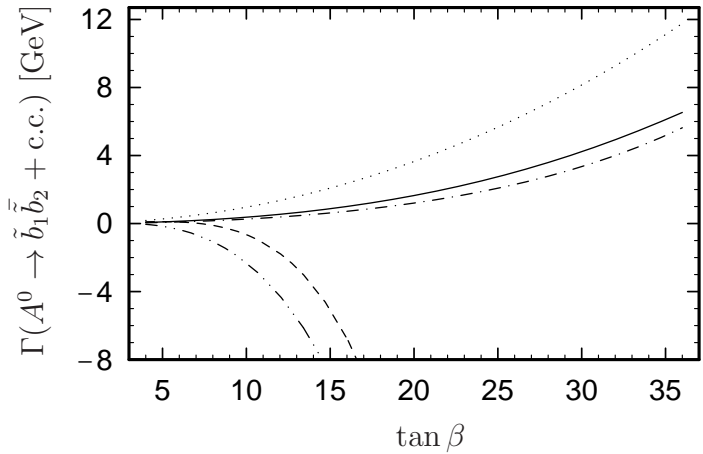}}}}
    \put(76,3){\makebox(0,0)[t]{{$\tan\b$}}}
    \put(20,12){\rotatebox{90}{{{$\Gamma (A^0 \rightarrow
                        \sb_1 \!\bar{\,\sb_2}+{\rm c.c.})$ [GeV]}}}}
\end{picture}
\caption{Two kinds of perturbation expansion: the dotted line 
corresponds to the on--shell tree--level width, the dashed and 
dash--dot--dotted line correspond to electroweak SUSY--QCD 
on--shell one--loop width, respectively. The dash--dotted line 
corresponds to improved the tree--level and the solid line to the 
(full) improved one--loop width.\label{tanbdep_sb_imp}} 
\end{figure}

In Fig.~\ref{tanbdep_sb_imp} we show two kinds of perturbation 
expansion for $\Gamma (A^0 \rightarrow \sb_1 \!\bar{\,\sb_2})$ 
with $\{m_{A^0}, \mu,$ $A, M, M_{\ti Q} \} = \{800, -300, -500, 
200, 300\}~\mbox{GeV}$: First we show the tree--level width, given 
in terms of on--shell input parameters (dotted line). The dashed 
and dash--dot--dotted line correspond to the on--shell electroweak 
and full (electroweak plus SUSY--QCD) one--loop width, 
respectively. For both corrections one can clearly see the 
invalidity of the on--shell perturbation expansion, which leads to 
an improper negative decay width. The second way of perturbation 
expansion is given by the dash--dotted and the solid line 
which correspond to the improved tree--level and improved full 
one--loop decay width, respectively. Here we take the same input 
parameters as in the first case but with running $A = 
-500~\mbox{GeV}$.

\begin{figure}[h!]
\begin{picture}(160,65)(0,0)
    \put(31,5){\mbox{\resizebox{8.7cm}{!}
    {\includegraphics{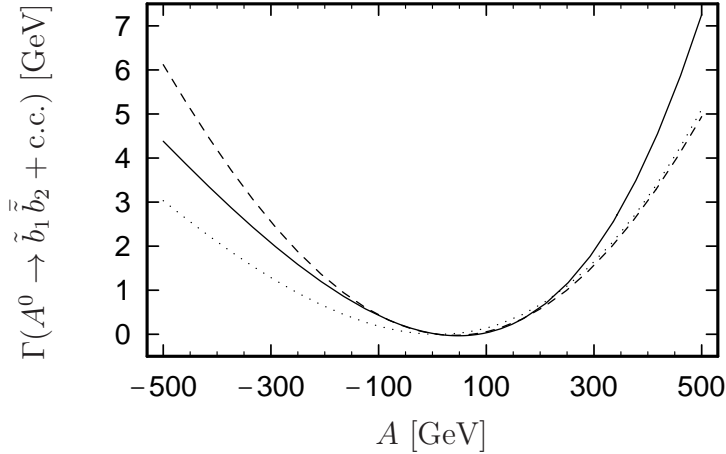}}}}
    \put(76,3){\makebox(0,0)[t]{{$A$ [GeV]}}}
    \put(20,11){\rotatebox{90}{{{$\Gamma (A^0 \rightarrow
                        \sb_1 \!\bar{\ \sb_2}+{\rm c.c.})$ [GeV]}}}}
\end{picture}
\caption{Perturbation expansion around the improved tree--level 
decay width (dotted line) of $\Gamma (A^0 \rightarrow \sb_1 
\!\bar{\ \sb_2})$ as a function of the trilinear coupling $A$. 
Dashed and solid lines correspond to the improved SUSY--QCD and 
full improved one--loop width, respectively.\label{Adep_sb_imp}} 
\end{figure}

In Fig.~\ref{Adep_sb_imp} we show the decay width $\Gamma (A^0 
\rightarrow \sb_1 \!\bar{\ \sb_2})$ as a function of 
$\overline{\rm DR}$ running $A$ for $\{m_{A^0}, \mu, M, M_{\ti Q} 
\} = \{800, -300, 300, 300\}~\mbox{GeV}$ and $\tan\b = 30$. The 
dotted line corresponds to the improved tree--level, the dashed 
line corresponds to the improved SUSY--QCD one--loop width and the 
solid line shows the full improved one--loop width. For negative 
$A$ the electroweak corrections decrease the decay width by $\sim 
20\%$, whereas for positive $A$ the SUSY--QCD corrections almost 
vanish and the electroweak ones go up to $30 \%$. 

\begin{figure}[h!]
\begin{picture}(160,58)(0,0)
    \put(28.5,5){\mbox{\resizebox{9cm}{!}
    {\includegraphics{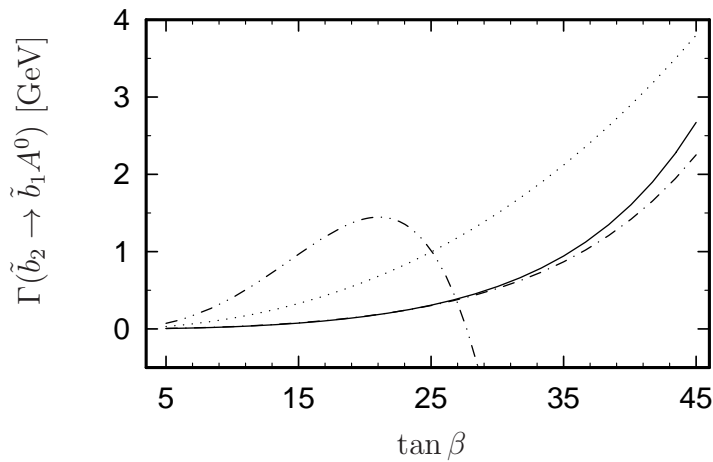}}}}
    \put(76,3){\makebox(0,0)[t]{{$\tan\b$}}}
    \put(20,20){\rotatebox{90}{{{$\Gamma (\sb_2 \rightarrow
                                  \sb_1 A^0)$ [GeV]}}}}
\end{picture}
\caption{$\tan\b$--dependence of $\Gamma (\sb_2 \rightarrow \sb_1 
A^0)$ for two kinds of perturbation expansion. The dotted and 
dash--dot--dotted lines corresponds to on--shell tree--level and 
full one--loop width, respectively, the dash--dotted line 
corresponds to improved tree--level and the solid line shows the 
full improved one--loop width. \label{tanbdep_sb_imp_crossed}} 
\end{figure}
Fig.~\ref{tanbdep_sb_imp_crossed} shows the behaviour of the decay 
width $\Gamma (\sb_2 \rightarrow \sb_1 A^0)$ for large $\tan\b$. 
As in Fig.~\ref{tanbdep_sb_imp} two kinds of perturbation 
expansion are given. The dotted and dash--dot--dotted lines 
correspond to the tree--level and full one--loop decay widths in 
the pure on--shell scheme. For large $\tan\b$ one can clearly see 
the invalidity of the perturbation series, leading to a negative 
decay width. In the second case we show the expansion around the 
tree--level decay width, given in terms of running $A_b$ and 
$m_b$. The dash--dotted line corresponds to the improved 
tree--level and the solid one to the one--loop width. Up to 
$\tan\b \sim 30$ the corrections stay relatively small which 
indicates that already the (improved) tree--level is a good 
approximation for $\Gamma (\sb_2 \rightarrow \sb_1 A^0)$. As input 
parameters we take the values $\{m_{A^0}, \mu, A, M, M_{\ti Q} \} 
= \{150, -220, 500, 200, 300\}~\mbox{GeV}$ and $M_{\ti D_3} = 
500~\mbox{GeV}$ for kinematical reasons.

\begin{figure}[h!]
\begin{picture}(160,65)(0,0)
    \put(28,5){\mbox{\resizebox{8.8cm}{!}
    {\includegraphics{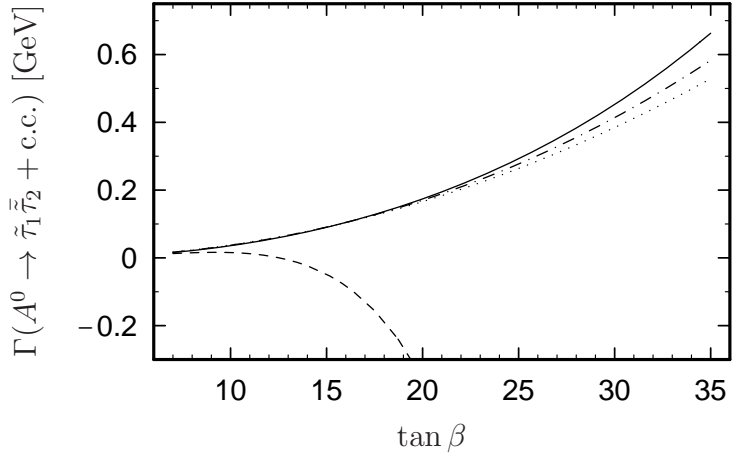}}}}
    \put(76,3){\makebox(0,0)[t]{{$\tan\b$}}}
    \put(20,12){\rotatebox{90}{{{$\Gamma (A^0 \rightarrow
                     \stau_1 \!\bar{\,\stau_2}+{\rm c.c.})$ [GeV]}}}}
\end{picture}
\caption{On--shell tree--level (dotted line) and full electroweak 
on--shell corrected decay width (dashed line) of $A^0 \rightarrow 
\stau_1 \!\bar{\,\stau_2}$ as a function of $\tan\b$. The 
dash--dotted and solid lines correspond to improved tree--level 
and full improved one--loop decay widths. 
\label{tanbdep_stau_imp}} 
\end{figure}

In Fig.~\ref{tanbdep_stau_imp} the $A^0$ decay into two staus is 
given as a function of $\tan\b$. Despite the absence of SUSY--QCD 
corrections the perturbation expansion around the on--shell 
tree--level (dotted line) leads to an improper negative decay 
width (dashed line) coming from large $\mathcal O (h_b^2)$ 
corrections. As input parameters we take $\{m_{A^0}, \mu, A, M, 
M_{\ti Q} \} = \{800, 400, -500, 120, 300\}~\mbox{GeV}$. The 
dash--dotted line corresponds to the improved tree--level and the 
solid line shows the improved one--loop width for the same input 
parameters as above and running $A = -500~\mbox{GeV}$. 

\begin{figure}[h!]
\begin{picture}(160,65)(0,0)
    \put(29,5){\mbox{\resizebox{8.7cm}{!}
    {\includegraphics{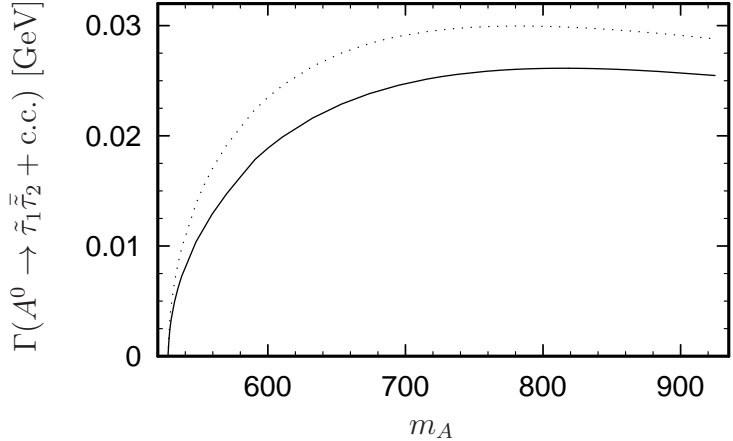}}}}
    \put(76,3){\makebox(0,0)[t]{{$m_A$}}}
    \put(20,12){\rotatebox{90}{{{$\Gamma (A^0 \rightarrow
                     \stau_1 \!\bar{\,\stau_2}+{\rm c.c.})$ [GeV]}}}}
\end{picture}
\caption{$m_{A^0}$--dependence of the improved tree--level (dotted 
line) and full one--loop corrected (solid line) decay widths of 
$\Gamma (A^0 \rightarrow \stau_1 \!\bar{\,\stau_2})$. 
\label{mAdep_stau_imp}} 
\end{figure}

Fig.~\ref{mAdep_stau_imp} shows the decay width of $\Gamma (A^0 
\rightarrow \stau_1 \!\bar{\,\stau_2})$ as a function of the mass 
of the decaying Higgs boson $A^0$ for the improved perturbation 
expansion. The dotted and the solid lines correspond to the 
(improved) tree--level and full one--loop widths, respectively. In 
the whole region of the parameter space shown the (electroweak) 
corrections decrease the on--shell width by 15\%. As input 
parameters we choose  $\{\mu, A, M, M_{\ti Q} \} = \{-450, -500, 
120, 260\}~\mbox{GeV}$ and $\tan\b = 7$. \clearpage

\begin{figure}[h!]
\begin{picture}(160,65)(0,0)
    \put(28.5,5){\mbox{\resizebox{8.7cm}{!}
    {\includegraphics{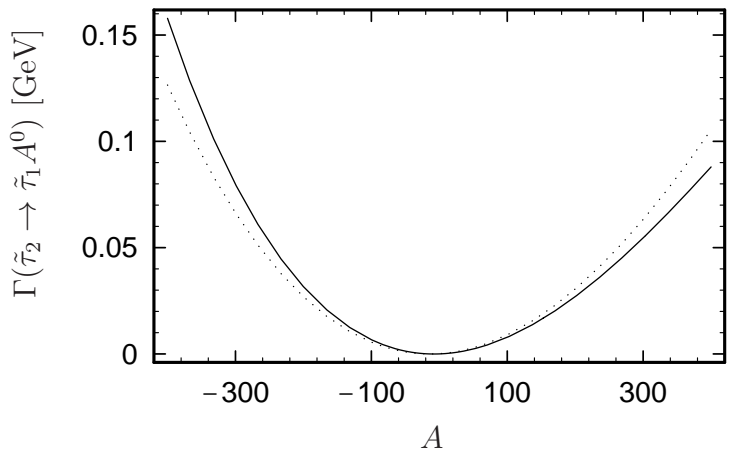}}}}
    \put(76,3){\makebox(0,0)[t]{{$A$}}}
    \put(20,20){\rotatebox{90}{{{$\Gamma (\stau_2 \rightarrow
                                  \stau_1 A^0)$ [GeV]}}}}
\end{picture}
\caption{$A$--dependence of the improved tree--level (dotted line) 
and improved one--loop decay width (solid line) for $\stau_2 
\rightarrow \stau_1 A^0$. \label{Adep_stau_imp_crossed}} 
\end{figure}

In Fig.~\ref{Adep_stau_imp_crossed} we show the $A$ dependence of 
$\Gamma (\stau_2 \rightarrow \stau_1 A^0)$ in the improved case. 
For negative values of $A$ the corrections increase the on--shell 
width by $\sim 20\%$ whereas for positive values of $A$ the 
corrections are negative and go up to $15 \%$. The input 
parameters are taken as follows: $\{m_{A^0}, \mu, M, M_{\ti Q} \} 
= \{150, 400, 300, 300\}~\mbox{GeV}$ and $M_{\ti E_3} = 
500~\mbox{GeV}$ for an acceptable stau mass splitting. For 
$\tan\b$ we take the value 30. 

\clearpage
\section{Conclusions}\label{conclusions}
\vspace{2mm} We have calculated the {\em full} electroweak 
one--loop corrections to the decay widths $A^0 \rightarrow \sf_1 \ 
{\bar{\!\sf_2}}$ and $\sf_2 \rightarrow \sf_1 A^0$ in the 
on--shell scheme. We have presented all formulae required for the 
computation. It has been necessary to renormalize almost all 
parameters of the MSSM. We have also included the SUSY--QCD 
corrections which were calculated in \cite{SUSY-QCD}. For the 
decay into bottom squarks and tau sleptons for large $\tan\b$ an 
improvement of the on--shell perturbation expansion is necessary. 
We have worked out an iterative method to improve the one--loop 
calculation. Thereby, the tree--level coupling is redefined in 
terms of $\overline{\rm DR}$ running masses and running $A_f$. We 
find that the corrections are significant and in a wide range of 
the parameter space comparable to the SUSY--QCD corrections.\\

\noindent {\bf Acknowledgements}\\ \noindent The authors
acknowledge support from EU under the HPRN-CT-2000-00149 network
programme and the ``Fonds zur F\"orderung der wissenschaftlichen
Forschung'' of Austria, project No. P13139-PHY.

\clearpage
\appendix

\section{Notation and Couplings}\label{appCouplings}
For the neutral and charged Higgs fields we use the notation
$H_k^0 = \{h^0, H^0, A^0, G^0\}$, $H_k^+ = \{H^+, G^+, H^-, G^-\}$
and $H_k^{-} \equiv (H_k^+)^\ast = \{H^-, G^-, H^+, G^+\}$. 
$t/\st$ stands for an up--type (s)fermion and $b/\sb$ for a 
down--type one. Following \cite{GunionHaber2, GunionHaber1} the 
Higgs--Sfermion--Sfermion couplings for neutral Higgs bosons, 
$G_{ijk}^\sf$, can be written as 
\begin{eqnarray}
G_{ijk}^\sf &\equiv& G\Big(H_k^0 \sf_i^\ast \sf_j\Big) = \Big[
R^\sf G_{LR, k}^\sf (R^\sf)^T \Big]_{ij} \,.
\end{eqnarray}
The 3rd generation left--right couplings $G_{LR, k}^\sf$ for up--
and down--type sfermions are
\begin{eqnarray}\non
G_{LR, 1}^\st &=& \left(\!
\begin{array}{cc}
-\sqrt 2 h_t m_t c_\a + g_{\scriptscriptstyle Z} 
m_{\scriptscriptstyle Z} (I_t^{3L}\!-\!e_t s_{\scriptscriptstyle 
W}^2) s_{\a+\b} & -\frac{h_t}{\sqrt 2} (A_t\, c_\a + \mu s_\a) \\ 
-\frac{h_t}{\sqrt 2} (A_t\, c_\a + \mu s_\a) & -\sqrt 2 h_t m_t 
c_\a + g_{\scriptscriptstyle Z} m_{\scriptscriptstyle Z} e_t 
s_{\scriptscriptstyle W}^2 s_{\a+\b} 
\end{array} \!\right) \,,
\\[2mm] \non
G_{LR, 1}^\sb &=& \left(\!
\begin{array}{cc}
\sqrt 2 h_b m_b s_\a + g_{\scriptscriptstyle Z} 
m_{\scriptscriptstyle Z} (I_b^{3L}\!-\!e_b s_{\scriptscriptstyle 
W}^2) s_{\a+\b} & \frac{h_b}{\sqrt 2} (A_b\, s_\a + \mu c_\a) \\ 
\frac{h_b}{\sqrt 2} (A_b\, s_\a + \mu c_\a) & \sqrt 2 h_b m_b s_\a 
+ g_{\scriptscriptstyle Z} m_{\scriptscriptstyle Z} e_b 
s_{\scriptscriptstyle W}^2 s_{\a+\b} 
\end{array} \!\right) \,,
\\[2mm] \non
&& G_{LR, 2}^\sf ~=~ G_{LR, 1}^\sf \qquad {\rm with}\ \a
\rightarrow \a - \pi/2 \,,
\\[2mm] \non
G_{LR, 3}^\st &=& -\sqrt 2 h_t \left(\!
\begin{array}{cc}
0 & -\frac{i}{2} \Big( A_t\, c_\b + \mu\, s_\b \Big) \\
\frac{i}{2} \Big( A_t\, c_\b + \mu\, s_\b \Big) & 0
\end{array} \!\right) \,,
\\[2mm] \non
G_{LR, 3}^\sb &=& -\sqrt 2 h_b \left(\! 
\begin{array}{cc}
0 & -\frac{i}{2} \Big( A_b\, s_\b + \mu\, c_\b \Big) \\
\frac{i}{2} \Big( A_b\, s_\b + \mu\, c_\b \Big) & 0
\end{array} \!\right) \,,
\\[2mm] \non
&& G_{LR, 4}^\sf ~=~ G_{LR, 3}^\sf \qquad {\rm with}\ \b
\rightarrow \b - \pi/2 \,,
\end{eqnarray}
where we have used the abbreviations $s_x \equiv \sin x$, $c_x
\equiv \cos x$ and $s_{\scriptscriptstyle W} \equiv \sin 
\theta_{\scriptscriptstyle W}$. $\a$ denotes the mixing angle of 
the $\{h^0, H^0\}$--system, and $h_t$ and $h_b$ are the Yukawa 
couplings 
\begin{eqnarray}
h_t ~=~ \frac{g\, m_t}{\sqrt 2 m_{\scriptscriptstyle W} \sin\b}\,, 
\qquad h_b ~=~ \frac{g\, m_b}{\sqrt 2 m_{\scriptscriptstyle W} 
\cos\b} \,. 
\end{eqnarray}
The couplings of charged Higgs bosons to two sfermions are given 
by $(l = 1, 2 )$ 
\begin{eqnarray}
G_{ijl}^{\sf\sf'} &\equiv& G\Big(H_l^\pm \sf_i^\ast \sf'_j\Big)
~=~ G_{jil}^{\sf'\!\sf} ~=~ \left( R^\sf \, G_{LR, l}^{\sf\sf'}
\left( R^{\sf'} \right)^T \right)_{ij} \,,
\end{eqnarray}
\begin{eqnarray}
G_{LR, 1}^{\st\sb} &=& \left(
    \begin{array}{cc}
        h_b m_b \sin\b + h_t m_t \cos\b - 
        \frac{g m_{\scriptscriptstyle W}}{\sqrt 2} \sin 2\b 
        & h_b (A_b \sin\b + \mu \cos\b)
        \\[2mm]
        h_t (A_t \cos\b + \mu \sin\b) & h_t m_b \cos\b + h_b m_t
        \sin\b
    \end{array}
\right) \,,
\\[4mm]\non
G_{LR, 1}^{\sb\st} &=& \left(
    \begin{array}{cc}
        h_b m_b \sin\b + h_t m_t \cos\b - 
        \frac{g m_{\scriptscriptstyle W}}{\sqrt 2} \sin 2\b
        & h_t (A_t \cos\b + \mu \sin\b)
        \\[2mm]
        h_b (A_b \sin\b + \mu \cos\b) & h_t m_b \cos\b + h_b m_t
        \sin\b
    \end{array}
\right) ~=~ \left( G_{LR, 1}^{\st\sb} \right)^T \,,
\\
\\
G_{LR, 2}^{\sf\sf'} &=& G_{LR, 1}^{\sf\sf'} \qquad {\textrm{with}}
\quad \b ~\rightarrow~ \b - \frac{\pi}{2} \,.
\end{eqnarray}
$f'$ denotes the isospin partner of the fermion $f$, i.~e. $t' = 
b,\, \sb_i' = \st_i$ etc. Note that only the angle $\b$ explicitly 
given in the matrices above has to be substituted; the dependence 
of $\b$ in the Yukawa couplings has to remain the same. 
\\ 

\noindent The $H_k^0 H_l^0 \sf_i^\ast \sf_j$ interaction is given
by
\begin{equation}
{\mathcal L} = - \frac{1}{2} \sum_f \Big[\, h_f^2\, c_{kl}^\sf\,
\d_{ij} + g^2\, \Big( c_{kl}^\sb - c_{kl}^\st \Big) e^\sf_{ij}
\,\Big]\, H^0_k H^0_l \sf_i^* \sf_j \,,
\end{equation}
with
\begin{eqnarray}
c_{kl}^\sb & = & \left(
    \begin{array}{cccc}
        \sin^2\!\alpha & - \frac{1}{2} \sin2\alpha & 0 & 0
        \\
        - \frac{1}{2} \sin2\alpha & \cos^2\!\alpha & 0 & 0
        \\
        0 & 0 & \sin^2\!\beta & - \frac{1}{2} \sin2\beta
        \\
        0 & 0 & - \frac{1}{2} \sin2\beta & \cos^2\!\beta
    \end{array}
\right)\,,
\\[3mm]
c_{kl}^\st & = & \left(
    \begin{array}{cccc}
        \cos^2\!\alpha &  \frac{1}{2} \sin2\alpha & 0 & 0
        \\
        \frac{1}{2} \sin2\alpha & \sin^2\!\alpha & 0 & 0
        \\
        0 & 0 & \cos^2\!\beta & \frac{1}{2} \sin2\beta
        \\
        0 & 0 & \frac{1}{2} \sin2\beta & \sin^2\!\beta
    \end{array}
\right)\,,
\\[3mm]
e_{ij}^\sf & = & \frac{1}{2 c_{\scriptscriptstyle W}^2} \bigg[ 
(I_f^{3L} - e_f s_{\scriptscriptstyle W}^2) R_{i1}^\sf R_{j1}^\sf 
+ e_f s_{\scriptscriptstyle W}^2 R_{i2}^\sf R_{j2}^\sf \bigg] \,. 
\end{eqnarray}
For the $H_k^+ H_l^- \sf_i^\ast \sf_j$ interaction,
\begin{eqnarray}
{\mathcal L} = - \frac{1}{2} \sum_f \Big[\, h_f^2\, d_{kl}^\sf
\Big( R^\sf_{i2} R^\sf_{j2} + R^{\sf'}_{i1} R^{\sf'}_{j1} \Big) +
g^2\, \Big( d_{kl}^\sb - d_{kl}^\st \Big) f^\sf_{ij} \,\Big]\,
H^+_k H^-_l \sf_i^* \sf_j \,,
\end{eqnarray}
we use the coupling matrices
\begin{eqnarray}
d_{kl}^\sb & = & \left(
    \begin{array}{cccc}
        \sin^2\!\beta & - \frac{1}{2} \sin2\beta & 0 & 0
        \\
        - \frac{1}{2} \sin2\beta & \cos^2\!\beta & 0 & 0
        \\
        0 & 0 & \sin^2\!\beta & - \frac{1}{2} \sin2\beta
        \\
        0 & 0 & - \frac{1}{2} \sin2\beta & \cos^2\!\beta
    \end{array}
\right)\,,
\\[3mm]
d_{kl}^\st & = & \left(
    \begin{array}{cccc}
        \cos^2\!\beta &  \frac{1}{2} \sin2\beta & 0 & 0
        \\
        \frac{1}{2} \sin2\beta & \sin^2\!\beta & 0 & 0
        \\
        0 & 0 & \cos^2\!\beta & \frac{1}{2} \sin2\beta
        \\
        0 & 0 & \frac{1}{2} \sin2\beta & \sin^2\!\beta
    \end{array}
\right)\,,
\\[3mm]
f^\sf_{ij} & = & \frac{1}{2 c_{\scriptscriptstyle W}^2} \bigg[ 
\big( - I_f^{3L} \cos 2\theta_{\scriptscriptstyle W} - e_f 
s_{\scriptscriptstyle W}^2 \big)\, R_{i1}^\sf R_{j1}^\sf + e_f 
s_{\scriptscriptstyle W}^2 R_{i2}^\sf R_{j2}^\sf \bigg] \,. 
\end{eqnarray}

\noindent For the Higgs--fermion--fermion couplings the
interaction Lagrangian reads
\begin{equation}
\mathcal{L} = \sum_{k=1}^2 s_k^f\, H_k^0 \bar f f + \sum_{k=3}^4 
s_k^f\, H_k^0 \bar f \gamma^5 f + \sum_{l=1}^2 \Big[ H_l^+ \bar t 
\, \big(y_l^b P_R \!+\! y_l^t P_L \big) \, b + {\rm h.c.} \Big] 
\end{equation}
with the couplings
\begin{equation}
    \begin{array}{ll}
        s_1^t ~=~ - g \frac{m_t \cos\alpha}{2m_{\scriptscriptstyle W}\sin\beta} ~=~
        -\frac{h_t}{\sqrt{2}}\cos\alpha,  \qquad\qquad & s_1^b ~=~ g\frac{m_b
        \sin\alpha}{2m_{\scriptscriptstyle W}\cos\beta} ~=~
        \frac{h_b}{\sqrt{2}}\sin\alpha \,, \non
        \\[5mm]
        s_2^t ~=~ - g \frac{m_t \sin\alpha}{2m_{\scriptscriptstyle W}\sin\beta} ~=~
        -\frac{h_t}{\sqrt{2}}\sin\alpha, \qquad\qquad & s_2^b ~=~ -g\frac{m_b
        \cos\alpha}{2m_{\scriptscriptstyle W}\cos\beta} ~=~
        -\frac{h_b}{\sqrt{2}}\cos\alpha \,, \non
        \\[5mm]
        s_3^t ~=~ ig\frac{m_t\cot\beta}{2m_{\scriptscriptstyle W}} ~=~
        i\frac{h_t}{\sqrt{2}}\cos\beta,  \qquad\qquad & s_3^b ~=~
        ig\frac{m_b\tan\beta}{2m_{\scriptscriptstyle W}} ~=~
        i\frac{h_b}{\sqrt{2}}\sin\beta \,, \non
        \\[5mm]
        s_4^t ~=~ ig\frac{m_t}{2m_{\scriptscriptstyle W}} ~=~
        i\frac{h_t}{\sqrt{2}}\sin\beta,  \qquad\qquad & s_4^b ~=~
        -ig\frac{m_b}{2m_{\scriptscriptstyle W}} ~=~
        -i\frac{h_b}{\sqrt{2}}\cos\beta \,, \non
        \\[5mm]
        y_1^t ~=~ g\frac{m_t\cot\beta}{\sqrt2 m_{\scriptscriptstyle W}} ~=~
        h_t \cos\beta, \qquad\qquad & y_1^b ~=~
        g\frac{m_b\tan\beta}{\sqrt2 m_{\scriptscriptstyle W}} ~=~
        h_b \sin\beta \,, \non
        \\[5mm]
        y_2^t ~=~ g\frac{m_t}{\sqrt2 m_{\scriptscriptstyle W}} ~=~
        h_t \sin\beta, \qquad\qquad & y_2^b ~=~
        -g\frac{m_b}{\sqrt2 m_{\scriptscriptstyle W}} ~=~
        -h_b \cos\beta \,. \non
    \end{array}
\end{equation}

\noindent The interaction Lagrangian for Higgs bosons and gauginos
is given by
\begin{eqnarray}\non
\mathcal L &=& - \frac{g}{2} \sum_{k=1}^2 H_k^0\,
{\bar{\tilde\chi}}_l^0 F_{lmk}^0 \, \nt_m - i \frac{g}{2}
\sum_{k=3}^4 H_k^0\, {\bar{\tilde\chi}}_l^0\, F_{lmk}^0 \g_5\,
\nt_m
\\[2mm] \non
&& - g \sum_{k=1}^2 H_k^0\, {\bar{\tilde\chi}}_i^+ \left(
F^+_{ijk} P_R + F^+_{jik} P_L \right) \chp_j + i g \sum_{k=3}^4
H_k^0\, {\bar{\tilde\chi}}_i^+ \left( F^+_{ijk} P_R + F^+_{jik}
P_L \right) \chp_j
\\[2mm]
&& - g \sum_{k=1}^2 \Big[ H_k^+ \, {\bar{\tilde\chi}}_i^+ \left(
F_{ilk}^R P_R + F_{ilk}^L P_L \right) \nt_l + {\rm h.c.} \Big] \,. 
\end{eqnarray}
with
\begin{eqnarray} \non
F_{lmk}^0 &=& \hphantom{+}\frac{e_k}{2} \Big[ Z_{l3} Z_{m2} +
Z_{m3} Z_{l2} - \tan\theta_{\scriptscriptstyle W} \left( Z_{l3} 
Z_{m1} + Z_{m3} Z_{l1} \right) \Big] 
\\[2mm]
&& + \frac{d_k}{2} \Big[ Z_{l4} Z_{m2} + Z_{m4} Z_{l2} -
\tan\theta_{\scriptscriptstyle W} \left( Z_{l4} Z_{m1} + Z_{m4} 
Z_{l1} \right) \Big] \ = \ F_{mlk}^0 \,, 
\\[2mm]
F_{ijk}^+ &=& \frac{1}{\sqrt2} \left( e_k V_{i1} U_{j2} - d_k
V_{i2} U_{j1} \right) \,,
\end{eqnarray}
and
\begin{eqnarray} \non
F_{ilk}^R &=& d_{k+2} \left[ V_{i1} Z_{l4} + \frac{1}{\sqrt2}
(Z_{l2} + Z_{l1} \tan\theta_{\scriptscriptstyle W}) V_{i2} \right] 
\,, 
\\[2mm]
F_{ilk}^L &=& -e_{k+2} \left[ U_{i1} Z_{l3} - \frac{1}{\sqrt2}
(Z_{l2} + Z_{l1} \tan\theta_{\scriptscriptstyle W}) U_{i2} \right] 
\,. 
\end{eqnarray}
$U, V$ and $Z$ are rotation matrices which diagonalize the
chargino and neutralino mass matrices. $d_k$ and $e_k$ take the
values
\begin{eqnarray}\non
d_k = \{ -\cos\a, -\sin\a, \cos\b, \sin\b \}\,, \qquad e_k = \{
-\sin\a, \cos\a, -\sin\b, \cos\b \} \,.
\end{eqnarray}

\noindent The coupling of the vector boson $Z^0$ to two sfermions, 
$\mathcal L = -i\, g_{\scriptscriptstyle Z}\, z^\sf_{ij}\, 
Z_\mu^0\, \sf_i^\ast 
\!\stackrel{\leftrightarrow}{\partial^\mu}\!\sf_j$ with 
$g_{\scriptscriptstyle Z}=g/\cos{\tw}$, is given by the matrix 
\begin{eqnarray}\non
z^\sf_{ij} &=& C^f_L \, R^\sf_{i1} R^\sf_{j1} + C^f_R \,
R^\sf_{i2} R^\sf_{j2}
\\[-3mm]\non
\end{eqnarray}
with $C^f_L = I_f^{3L} - e_f s_{\scriptscriptstyle W}^2$ and
$C^f_R = - e_f s_{\scriptscriptstyle W}^2$.

\noindent For the interaction of a vector boson with two gauginos 
we use the couplings
\begin{eqnarray}\non
& O_{ij}^{L} ~=~ Z_{i2} V_{j1} - \frac{1}{\sqrt{2}} Z_{i4} V_{j2}
\,, \qquad O_{ij}^{R} ~=~ Z_{i2} U_{j1} + \frac{1}{\sqrt{2}}
Z_{i3} U_{j2} \,, &
\\[3mm]\non
& O_{ij}^{'L} ~=~ - V_{i1} V_{j1} - \frac{1}{2} V_{i2} V_{j2} +
\delta_{ij} s_{\scriptscriptstyle W}^2 \,, &
\\[2mm]\non
& O_{ij}^{'R} ~=~ - U_{i1} U_{j1} - \frac{1}{2} U_{i2} U_{j2} +
\delta_{ij} s_{\scriptscriptstyle W}^2 \,, &
\\[3mm]\non
& O_{ij}^{''L} ~=~ - \frac{1}{2} Z_{i3} Z_{j3} + \frac{1}{2}
Z_{i4} Z_{j4} ~=~ - O_{ij}^{''R} \,. &
\end{eqnarray}

\noindent The interaction Lagrangian of the
chargino--sfermion--fermion couplings is given by
\begin{eqnarray}\non
        \mathcal L &=& \bar t \left(l_{ij}^\sb P_R + k_{ij}^\sb P_L\right)
        \chp_j \sb_i + \bar b \left( l_{ij}^\st P_R + k_{ij}^\st P_L
        \right) \tilde\chi^{+c}_j \,\st_i \vspace{4pt}
        \\[2mm]
        && +~ \overline{\chp_j} \left(l_{ij}^\sb P_L + k_{ij}^\sb
        P_R \right) \!t\, {\sb_i}^{\ast} + \overline{\tilde\chi^{+c}_j}
        \left(l_{ij}^\st P_L + k_{ij}^\st P_R \right)\! b\,
        \st_i^\ast
\end{eqnarray}
with the coupling matrices
\begin{equation}
\begin{array}{rcl@{\qquad\qquad}lcl}
    l_{ij}^\st &=& -g V_{j1} R_{i1}^\st + h_t V_{j2} R_{i2}^\st \,,
    & l_{ij}^\sb &=& -g U_{j1} R_{i1}^\sb + h_b U_{j2} R_{i2}^\sb\,,
    \\[4mm]
    k_{ij}^\st &=& h_b U_{j2} R_{i1}^\st\,, & k_{ij}^\sb &=& h_t
    V_{j2} R_{i1}^\sb\,.
\end{array}
\end{equation}
For the neutralino--sfermion--fermion couplings the Lagrangian
reads
\begin{eqnarray}
\mathcal L & = & \bar f \left(a_{ik}^\sf P_R + b_{ik}^\sf
P_L\right) \nt_k\,\sf_i + \bar{\ti \x}^0_k \left(a_{ik}^\sf P_L +
b_{ik}^\sf P_R\right) f \sf_i^\ast
\end{eqnarray}
with the coupling matrices
\begin{eqnarray}
a_{ik}^\sf &=& h_f Z_{kx} R_{i2}^\sf + g f_{Lk}^f R_{i1}^\sf \,,
\hspace{18mm} b_{ik}^\sf ~=~ h_f Z_{kx} R_{i1}^\sf + g f_{Rk}^f
R_{i2}^\sf
\end{eqnarray}
and
\begin{eqnarray}
f_{Lk}^f = \sqrt2 \left( 
(e_f-I_f^{3L})\tan\theta_{\scriptscriptstyle W} Z_{k1} + I_f^{3L} 
Z_{k2} \right)\,,\quad f_{Rk}^f = -\sqrt2 
e_f\tan\theta_{\scriptscriptstyle W} Z_{k1} \,. 
\end{eqnarray}
$x$ takes the values $\{3, 4\}$ for $\{$down, up$\}$--type case,
respectively.

\section{Vertex corrections}\label{appVertex}
\begin{figure}[th]
\begin{picture}(160,215)(0,0)
     \put(0,-2){\mbox{\resizebox{16cm}{!}
     {\includegraphics{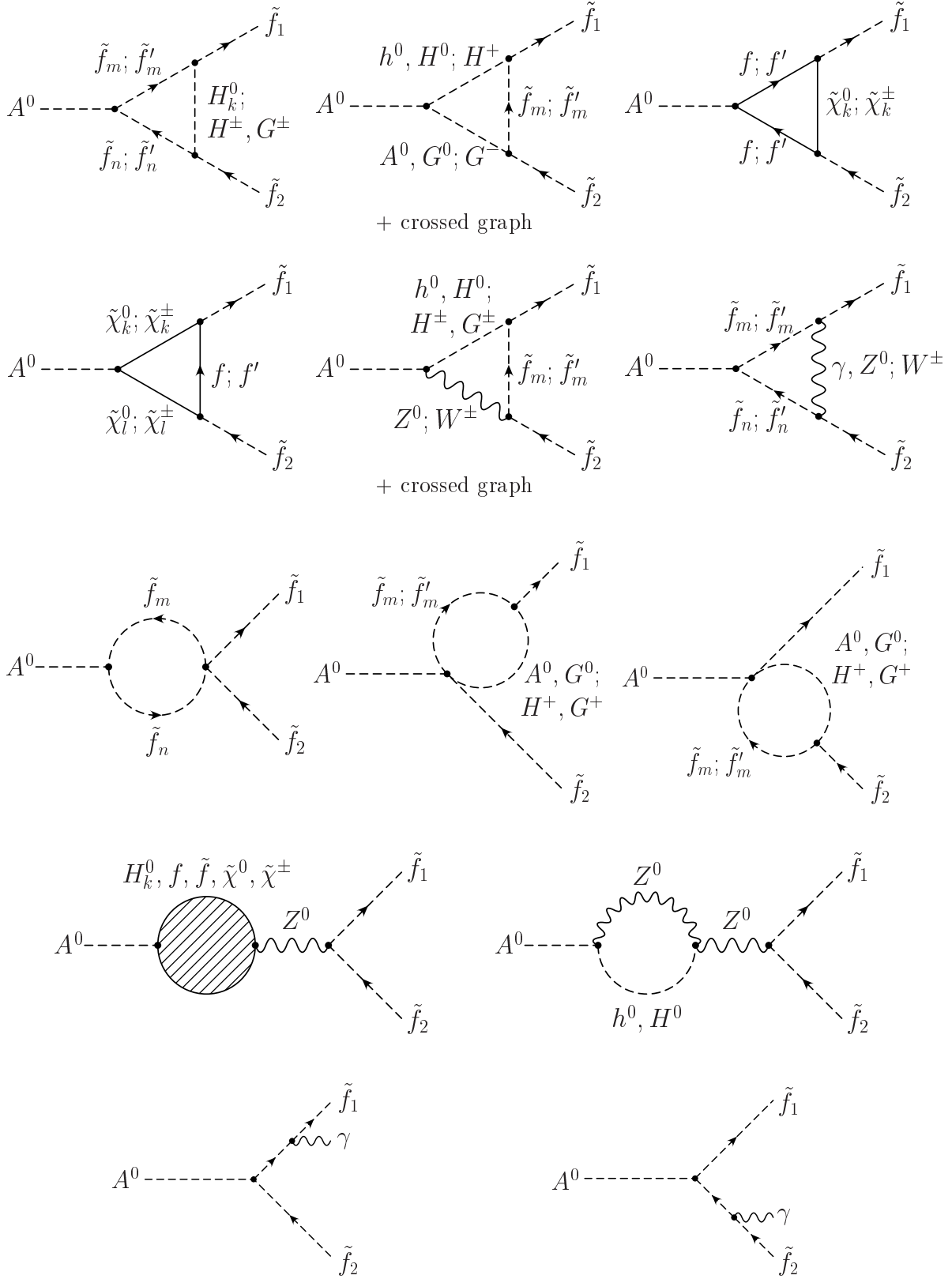}}}}
\end{picture}
\caption{Vertex and photon emission diagrams relevant to the 
calculation of the virtual electroweak corrections to the decay 
width $A^0 \rightarrow \tilde{f}_1 \ {\bar{\!\!\tilde{f}}}_{\!2}$. 
\label{vertex-graphs}} 
\end{figure}

Here we give the explicit form of the electroweak contributions to
the vertex corrections which are depicted in
Fig.~\ref{vertex-graphs}. For SUSY--QCD contributions we refer to
\cite{SUSY-QCD}.
\begin{eqnarray}
   \d G_{123}^{\sf (v)} &=& \d G_{123}^{\sf (v, H\sf\sf)}
      + \d G_{123}^{\sf (v, \sf HH)}
      + \d G_{123}^{\sf (v, \ti\chi ff)}
      + \d G_{123}^{\sf (v, f \ti\chi \ti\chi)}
   \non \\[3mm]
   && +~\d G_{123}^{\sf (v, V)}
      + \d G_{123}^{\sf (v, \sf\sf)}
      + \d G_{123}^{\sf (v, H\sf)}
      + \d G_{123}^{\sf (v, AZ)}
      + \d G_{123}^{\sf (v, AG)}
\end{eqnarray}
\begin{spacing}{1.8}
\noindent The single contributions correspond to the diagrams with
three scalar particles $\Big( \d G_{123}^{\sf (v, H\sf\sf)}$ and
$\d G_{123}^{\sf (v, HH\sf)} \Big)$, three fermions $\Big( \d
G_{123}^{\sf (v, \ti\chi ff)}$ and $\d G_{123}^{\sf (v, f \ti\chi
\ti\chi)} \Big)$, one vector particle $\Big( \d G_{123}^{\sf (v,
V)} \Big)$ or two scalar particles $\Big( \d G_{123}^{\sf (v,
\sf\sf)}$ and $\d G_{123}^{\sf (v, H\sf)} \Big)$ in the loop. $\d
G_{123}^{\sf (v, AZ\rm mix)}$ denotes the correction due to the
mixing of $A^0$ and $Z^0$ and $\d G_{123}^{\sf (v, AG)}$ is the
Higgs mixing transition $A^0$--$G^0$.
\end{spacing} 
\noindent As shown in \cite{0305250} we can sum up the $A^0 Z^0$ 
and $A^0 G^0$ transition amplitudes which leads to 
\begin{eqnarray}
\d G_{123}^{\sf (v, AZ)} + \d G_{123}^{\sf (v, AG)} &=& -
\frac{i}{m_{\scriptscriptstyle Z}}\, \Pi_{AZ}(m_{A^0}^2)\,
G^\sf_{124} \,.
\end{eqnarray}
The explicit form of the $A^0$--$Z^0$ self--energy,
$\Pi_{AZ}(m_{A^0}^2)$, is given in app.~\ref{appAZmixing}.
\noindent The vertex corrections from the exchange of one Higgs
and two sfermions are
\begin{eqnarray}\non
\d G_{123}^{\sf (v, H\sf\sf)} &=& - \frac{1}{(4\pi)^2} \sum_{m,n =
1}^2 \sum_{k=1}^4 G_{mn3}^\sf G_{imk}^\sf G_{njk}^\sf\, C_0\Big(
m_{\sf_i}^2, m_{A^0}^2, m_{\sf_j}^2, m_{H_k^0}^2, m_{\sf_m}^2,
m_{\sf_n}^2 \Big)
\\[2mm] \non
&& - \frac{1}{(4\pi)^2} \sum_{m,n = 1}^2 \sum_{k=1}^2
G_{mn3}^{\sf'} G_{imk}^{\sf\sf'} G_{jnk}^{\sf\sf'}\, C_0\Big(
m_{\sf_i}^2, m_{A^0}^2, m_{\sf_j}^2, m_{H_k^+}^2, m_{\sf'_m}^2,
m_{\sf'_n}^2 \Big)
\\[-3mm]
\\[-7mm] \non
\end{eqnarray}
with the standard two--point function $C_0$ \cite{PaVe} for which
we follow the conventions of \cite{Denner}. The graph with 2 Higgs
particles and one sfermion in the loop leads to
\begin{eqnarray}\non
\d G_{123}^{\sf (v, \sf HH)} &=& - \frac{1}{(4\pi)^2} 
\frac{g_{\scriptscriptstyle Z} \, m_{\scriptscriptstyle 
Z}}{2}\sum_{m = 1}^2 \left( \, \sum_{k = 1}^2 \sum_{l = 3}^4 
G^\sf_{imk} G^\sf_{mjl} \, A_{k,l-2} + \sum_{k = 3}^4 \sum_{l = 
1}^2 G^\sf_{imk} G^\sf_{mjl} \, A_{l,k-2} \right) \, \times 
\\[2mm]\non
&& \hspace{51mm} C_0\Big( m_{\sf_i}^2, m_{A^0}^2, m_{\sf_j}^2,
m_{\sf_m}^2, m_{H_k^0}^2, m_{H_l^0}^2 \Big)
\\[2mm]\non
&& - \frac{i}{(4\pi)^2} \, I_f^{3L} \, g \, m_{\scriptscriptstyle
W} \sum_{m = 1}^2 \bigg( G^{\sf\sf'}_{im1} G^{\sf\sf'}_{jm2}\,
C_0\Big( m_{\sf_i}^2, m_{A^0}^2, m_{\sf_j}^2, m_{\sf'_m}^2,
m_{H^+}^2, m_{G^+}^2 \Big)
\\[2mm]\non
&& \hspace{40mm} -G^{\sf\sf'}_{im2}\, G^{\sf\sf'}_{jm1}\, C_0\Big(
m_{\sf_i}^2, m_{A^0}^2, m_{\sf_j}^2, m_{\sf'_m}^2, m_{G^+}^2,
m_{H^+}^2 \Big) \bigg)
\end{eqnarray}
with
\begin{eqnarray}\non
   A_{kl} &=& \left(
   \begin{array}{rr}
      -\cos 2\b \, \sin(\a+\b) & -\sin 2\b \, \sin(\a+\b)
      \\[1mm]
      \cos 2\b \, \cos(\a+\b) & \sin 2\b \, \cos(\a+\b)
   \end{array}
   \right)\,.
\end{eqnarray}
For the gaugino exchange contributions we get
\begin{eqnarray}\non
\d G_{123}^{\sf (v, \ti\chi ff)} &=& \frac{1}{(4\pi)^2} \sum_{k =
1}^4 F\Big( m_{\sf_i}^2, m_{A^0}^2, m_{\sf_j}^2, m_{\nt_k}, m_f,
m_f; s^f_3, -s^f_3, b^\sf_{ik}, a^\sf_{ik}, a^\sf_{jk}, b^\sf_{jk}
\Big)
\\[1mm]\non
&& \hspace{-3mm}+\frac{1}{(4\pi)^2} \sum_{k = 1}^2 F\Big(
m_{\sf_i}^2, m_{A^0}^2, m_{\sf_j}^2, m_{\chp_k}, m_{f'}, m_{f'};
s^{f'}_3, -s^{f'}_3, k^\sf_{ik}, l^\sf_{ik}, l^\sf_{jk},
k^\sf_{jk} \Big)\,,
\\[3mm]\non
\d G_{123}^{\sf (v, f \ti\chi \ti\chi)} &=& \frac{1}{(4\pi)^2}
\sum_{k,l = 1}^4 \!\! F\Big( m_{\sf_i}^2, m_{A^0}^2, m_{\sf_j}^2,
m_f, m_{\nt_k}, m_{\nt_l}; i g F^0_{lk3}, -i g F^0_{lk3},
b^\sf_{ik}, a^\sf_{ik}, a^\sf_{jl}, b^\sf_{jl} \Big)
\\[1mm]\non
&& \hspace{-3mm}+\frac{1}{(4\pi)^2} \sum_{k,l = 1}^2 \!\! F\Big(
m_{\sf_i}^2, m_{A^0}^2, m_{\sf_j}^2, m_{f'}, m_{\chp_k},
m_{\chp_l}; i g \widetilde F^+_{kl3}, -i g \widetilde F^+_{lk3},
k^\sf_{ik}, l^\sf_{ik}, l^\sf_{jl}, k^\sf_{jl} \Big)\,,
\\
\end{eqnarray}
where $F(\ldots)$ shortly stands for
\begin{eqnarray}\non
F\Big( m_1^2, m_0^2, m_2^2, M_0, M_1, M_2; g_0^R, g_0^L, g_1^R,
g_1^L, g_2^R, g_2^L \Big) &=& (h_1M_1 \!+\! h_2M_2)B_0(m_0^2,
M_1^2, M_2^2)
\\[2mm]\non
&&\hspace{-9cm} +~(h_0 M_0 \!+\! h_1 M_1)B_0(m_1^2, M_0^2, M_1^2)
+(h_0M_0 \!+\! h_2M_2)B_0(m_2^2, M_0^2, M_2^2)
\\[2mm]\non
&&\hspace{-9cm} +\, \Big[2\!\left(g_0^Rg_1^Rg_2^R \!+\!
g_0^Lg_1^Lg_2^L \right)\!M_0M_1M_2 + h_0M_0\!\left( M_1^2\!
+\!M_2^2\!- \!m_0^2 \right) +
h_1M_1\!\left(M_0^2\!+\!M_2^2\!-\!m_2^2\right)
\\[2mm]
&&\hspace{-9cm}
+~h_2M_2\!\left(M_0^2\!+\!M_1^2\!-\!m_1^2\right)\Big] C_0(m_1^2,
m_0^2, m_2^2, M_0^2, M_1^2, M_2^2)
\end{eqnarray}
with the abbreviations $h_0 = \left(g_0^Lg_1^Rg_2^R +
g_0^Rg_1^Lg_2^L \right), h_1 = \left(g_0^Lg_1^Lg_2^R +
g_0^Rg_1^Rg_2^L \right)$ and $h_2 = \left(g_0^Rg_1^Lg_2^R +
g_0^Lg_1^Rg_2^L \right)$. For up--type sfermions $\widetilde
F^+_{kl3} = F^+_{kl3}$ and for down--type sfermions chargino
indices are interchanged, $\widetilde F^+_{kl3} = F^+_{lk3}$\,.
\\
We split the irreducible vertex graphs with one vector particle in
the loop into the single contributions of the photon, the
$Z$--boson and the $W$--boson,
\begin{eqnarray}
   \d G_{123}^{\sf (v, V)} &=& \d G_{123}^{\sf (v, \g)}
   + \d G_{123}^{\sf (v, Z)} + \d G_{123}^{\sf (v, W)}\,.
\end{eqnarray}
In order to regularize the infrared divergences we introduce a
photon mass $\l$. Thus we have
\begin{eqnarray}\non
\d G_{123}^{\sf (v, \g)} &=& \frac{1}{(4\pi)^2}\, (e_0 e_f)^2
G^\sf_{123}\ V\Big( m_{\sf_i}^2, m_{A^0}^2, m_{\sf_j}^2, \l^2,
m_{\sf_i}^2, m_{\sf_j}^2 \Big) \,,
\\ \non
\d G_{123}^{\sf (v, Z)} &=& \frac{1}{(4\pi)^2}\, 
g_{\scriptscriptstyle Z}^2 \sum_{m,n = 1}^2 G^\sf_{mn3}\, 
z^\sf_{im}\, z^\sf_{nj}\ V\Big( m_{\sf_i}^2, m_{A^0}^2, 
m_{\sf_j}^2, m_{\scriptscriptstyle Z}^2, m_{\sf_m}^2, m_{\sf_n}^2 
\Big) 
\\ \non
&& \hspace{-3mm}-\frac{i}{(4\pi)^2}\, \frac{g_{\scriptscriptstyle 
Z}^2}{2} \sum_{k,m = 1}^2 G^\sf_{mjk}\, z^\sf_{im}\, 
R_{1k}(\a\!-\!\b)\ V\Big( m_{A^0}^2, m_{\sf_j}^2, m_{\sf_i}^2, 
m_{\scriptscriptstyle Z}^2, m_{H_k^0}^2, m_{\sf_m}^2 \Big) 
\\ \non
&& \hspace{-3mm}+\frac{i}{(4\pi)^2}\, \frac{g_{\scriptscriptstyle 
Z}^2}{2} \sum_{k,m = 1}^2 G^\sf_{imk}\, z^\sf_{mj}\, 
R_{1k}(\a\!-\!\b)\ V\Big( m_{\sf_j}^2, m_{\sf_i}^2, m_{A^0}^2, 
m_{\scriptscriptstyle Z}^2, m_{\sf_m}^2, m_{H_k^0}^2 \Big) \,, 
\\ \non
\d G_{123}^{\sf (v, W)} &=& \frac{1}{(4\pi)^2}\, \frac{g^2}{2}
\sum_{m,n = 1}^2 G^{\sf'}_{mn3}\, R^\sf_{i1} R^\sf_{j1}
R^{\sf'}_{m1} R^{\sf'}_{n1} \ V\Big( m_{\sf_i}^2, m_{A^0}^2,
m_{\sf_j}^2, m_{\scriptscriptstyle W}^2, m_{\sf'_m}^2,
m_{\sf'_n}^2 \Big)
\\ \non
&& \hspace{-3mm}+\frac{i}{(4\pi)^2}\, \frac{g^2}{2\sqrt2} \sum_{m
= 1}^2 G^{\sf\sf'}_{jm1}\, R^\sf_{i1} R^{\sf'}_{m1} \ V\Big(
m_{A^0}^2, m_{\sf_j}^2, m_{\sf_i}^2, m_{\scriptscriptstyle W}^2,
m_{H^+}^2, m_{\sf'_m}^2 \Big)
\\
&& \hspace{-3mm}-\frac{i}{(4\pi)^2}\, \frac{g^2}{2\sqrt2} \sum_{m
= 1}^2 G^{\sf\sf'}_{im1}\, R^{\sf'}_{m1} R^\sf_{j1} V\Big(
m_{\sf_j}^2, m_{\sf_i}^2, m_{A^0}^2, m_{\scriptscriptstyle W}^2,
m_{\sf'_m}^2, m_{H^+}^2 \Big) \,,
\end{eqnarray}
where we have used the vector vertex function 
\begin{eqnarray}\non
V\Big( m_1^2, m_0^2, m_2^2, M_0^2, M_1^2, M_2^2 \Big) \!\!&=&\!\!
- B_0\Big( m_0^2, M_1^2, M_2^2\Big) \!+\! B_0\Big( m_1^2, M_0^2,
M_1^2\Big) \!+\!B_0\Big( m_2^2, M_0^2, M_2^2\Big)
\\[1mm]
&&\hspace{-35mm} +~\left( -2m_0^2 \!+\! m_1^2 \!+\! m_2^2 \!-\!
M_0^2 \!+\! M_1^2 \!+\! M_2^2 \right) C_0\Big( m_1^2, m_0^2,
m_2^2, M_0^2, M_1^2, M_2^2 \Big)
\end{eqnarray}
and the rotation matrix $R_{kl}$,
\begin{eqnarray}\non
R_{kl}(\phi) & \equiv & \left(
      \begin{array}{rr}
         \cos\phi & \sin\phi
         \\
         -\sin\phi & \cos\phi
      \end{array} \right)_{kl}\,.
\end{eqnarray}
$z^\sf_{ij}$ can be found in Appendix \ref{appCouplings}. \\ For 
the vertex graphs with 2 sfermions in the loop we obtain 
\begin{eqnarray}\non
\d G_{123}^{\sf (v, \sf\sf)} \!&=&\!
-\frac{1}{(4\pi)^2}\,h_f^2\!\! \sum_{m,n=1}^2 \!\! G_{nm3}^\sf
\bigg[ R^\sf_{ijmn} \!+\! R^\sf_{mnij} \!+\! N_C^f \Big(
R^\sf_{inmj} \!+\! R^\sf_{mjin} \Big)\! \bigg] B_0 \Big(
m_{A^0}^2, m_{\sf_m}^2, m_{\sf_n}^2 \Big)
\\[1mm] \non
&& -\frac{1}{(4\pi)^2}\, g_{\scriptscriptstyle Z}^2 \!\! 
\sum_{m,n=1}^2 \!\! G_{nm3}^\sf \Bigg\{ \bigg[ \Big( \frac{1}{4} - 
(2 I_f^{3L} \!-\! e_f) e_f s_{\scriptscriptstyle W}^2 \Big) 
R^{\sf_L}_{ijmn} + e_f^2 s_{\scriptscriptstyle W}^2 
R^{\sf_R}_{ijmn} \bigg] (N_C^f + 1) 
\\ \non
&&\hspace{38mm} + (I_f^{3L} \!-\! e_f) e_f s_{\scriptscriptstyle
W}^2 \bigg[ N_C^f \Big( R^{\sf}_{ijmn} \!+\! R^{\sf}_{mnij} \Big)
\!+\! R^{\sf}_{inmj} \!+\! R^{\sf}_{mjin} \bigg] \! \Bigg\}
\\[1mm] \non
&&\hspace{38mm} \times \, B_0 \Big( m_{A^0}^2, m_{\sf_m}^2,
m_{\sf_n}^2 \Big)
\\[1mm] \label{vertex2sf}
&&-\frac{1}{(4\pi)^2}\, N_C^{\hat f} \,h_f h_{\hat
f}\sum_{m,n=1}^2 \!\! G_{nm3}^{\ \hat{\!\!\sf}} \Big( R^{\sf\
\!\hat{\!\!\sf}_{\!F}}_{ijnm} + R^{\sf\
\!\hat{\!\!\sf}_{\!F}}_{jimn} \Big) B_0 \Big( m_{A^0}^2, m_{\
\!\hat{\!\!\sf}_{\!m}}^2, m_{\ \!\hat{\!\!\sf}_{\!n}}^2 \Big) \,.
\end{eqnarray}
For various products of sfermion rotation matrices we have
introduced the short forms
\begin{equation}\label{4sfconvention}
\begin{array}{l@{\,, \qquad\qquad}l}
   R^{\sf_L}_{ijkl} ~=~ R^\sf_{i1} R^\sf_{j1} R^\sf_{k1} R^\sf_{l1} &
   R^{\sf}_{ijkl} ~=~ R^\sf_{i1} R^\sf_{j1} R^\sf_{k2} R^\sf_{l2}\,,
   \\[1mm]
   R^{\sf_R}_{ijkl} ~=~ R^\sf_{i2} R^\sf_{j2} R^\sf_{k2} R^\sf_{l2} &
   R^{\sf\ \!\hat{\!\!\sf}_{\!F}}_{ijkl} ~=~ R^\sf_{i1} R^\sf_{j2}
   R^{\ \!\hat{\!\!\sf}}_{k1} R^{\ \!\hat{\!\!\sf}}_{l2}\,.
\end{array}
\end{equation}
Note that the last term in eq.~(\ref{vertex2sf}) originates from
the mixing of 2 squarks and 2 sleptons, where $\hat f$ denotes the
'family partner` of the fermion $f$ with the same isospin and from
the same generation, i.e. $\hat t = \nu_\tau$ or $\ \hat{\stau_i}
= \sb_i$. The diagrams with one Higgs boson and one sfermion in
the loop lead to
\begin{eqnarray}\non
\d G_{123}^{\sf (v, H\sf)} &=& -\frac{1}{(4\pi)^2}\,
\sum_{k=3}^4\, \sum_{m=1}^2 G^\sf_{imk} \bigg( h_f^2\,
c^\sf_{3k}\, \d_{mj} + g^2 \Big( c^\sb_{3k} - c^\st_{3k} \Big)
e^\sf_{mj} \bigg) B_0 \Big( m_{A^0}^2, m_{H_k^0}^2, m_{\sf_m}^2
\Big)
\\[2mm] \non
&&+\frac{i}{(4\pi)^2}\,\sqrt2 \, I_f^{3L}\!\! \sum_{k,m=1}^2 \!
G^{\sf\sf'}_{imk} \!\left[ \left\{ \!\!
\begin{array}{ll}
(h_\uparrow^2 \!-\! g^2/2) \cos^2\b - (h_\downarrow^2 \!-\! g^2/2)
\sin^2\b
\\[2mm]
(h_f^2 + h_{f'}^2 - g^2) \sin\b \cos\b
\end{array} \!\right\}_{\!k} \!
R^{\sf'}_{m1} R^\sf_{j1} \right.
\\[2mm] \non
&&\hspace{62mm} + h_f h_{f'}\, \d_{k2}\, R^{\sf'}_{m2} R^\sf_{j2}
\Bigg] B_0 \Big( m_{A^0}^2, m_{H_k^+}^2, m_{\sf'_m}^2 \Big)
\\[2mm] \non
&&\ -\ i \leftrightarrow j \,.
\end{eqnarray}
with $h_\uparrow = \{h_t, 0\}$ and $h_\downarrow = \{h_b,
h_\tau\}$ for the decay into \{squarks, sleptons\}, respectively.
The Higgs--Sfermion coupling matrices $c^\sf_{kl}$ and
$e^\sf_{ij}$ can be found in Appendix {\ref{appCouplings}}.

\clearpage

\section{Diagonal Wave--function corrections}\label{appWF}
\begin{figure}[th]
\begin{picture}(170,105)(0,0)
     \put(0,0){\mbox{\resizebox{16cm}{!}
     {\includegraphics{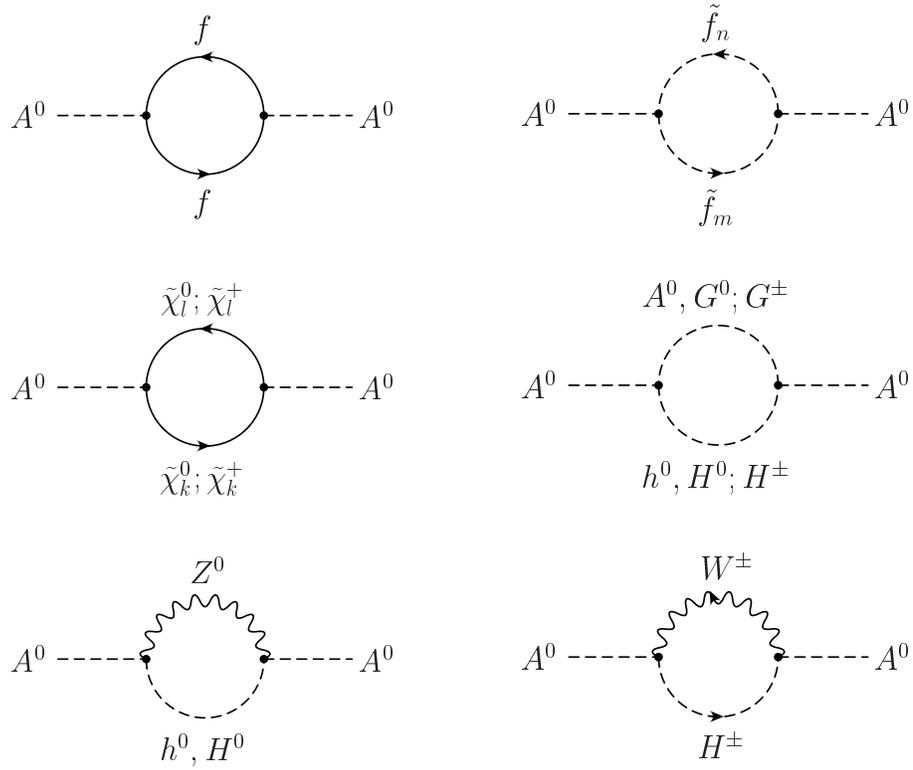}}}}
\end{picture}
\caption{Diagonal Higgs self--energies\label{A0SEdiag}}
\end{figure}

\begin{figure}[th]
\begin{picture}(170,70)(0,0)
     \put(0,0){\mbox{\resizebox{16cm}{!}
     {\includegraphics{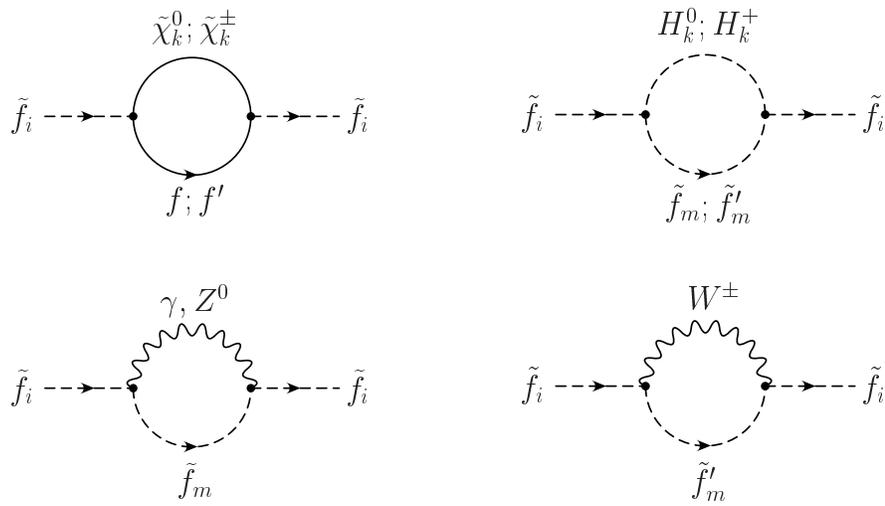}}}}
\end{picture}
\caption{Diagonal sfermion self--energies\label{SfermionSEdiag}}
\end{figure}

For the diagonal wave--function renormalization constants we use 
the conventional on--shell renormalization conditions which lead 
to 
\begin{eqnarray}
\delta Z_{33}^{H} ~=~ - \Re\,\dot\Pi_{33}^{H} (m_{A^0}^2) \,, 
\qquad \delta Z_{ii}^{\sf} ~=~ - \Re\,\dot\Pi_{ii}^{\sf} 
(m_{\sf_i}^2) \,,
\end{eqnarray}
where the dot in $\dot\Pi_{ii} (k^2)$ denotes the derivative with 
respect to $k^2$. In the following we list the single 
contributions of the wave--function corrections 
(Fig.~\ref{A0SEdiag} and \ref{SfermionSEdiag}).

\subsection{Higgs part}
\begin{eqnarray}
   \delta Z_{33}^{H,f} &=& \frac{2}{(4\pi)^2}\, \sum_{f}
   N_C^f \left(s^f_3\right)^2
   \bigg[ B_0( m_{A^0}^2, m_f^2, m_f^2 ) + m_{A^0}^2
   \dot{B}_0( m_{A^0}^2, m_f^2, m_f^2 )\bigg]
\\[2mm]
   \delta Z_{33}^{H,\sf} &=& -\frac{1}{(4\pi)^2}\, \sum_{f}
   \sum_{m,n=1}^2 N_C^f\, G^\sf_{mn3}\, G^\sf_{nm3}\,
   \dot{B}_0( m_{A^0}^2, m_{\sf_m}^2, m_{\sf_n}^2 )
\\[2mm]\non 
   \delta Z_{33}^{H,\nt} &=& \frac{1}{(4\pi)^2}\, g^2
   \sum_{k,l=1}^4 (F^0_{kl3})^2 \bigg[
   \dot{B}_0(m_{A^0}^2, m_{\nt_k}^2, m_{\nt_l}^2) \Big(
   (m_{\nt_k}-m_{\nt_l})^2 - m_{A^0}^2 \Big)
\\
&&\hspace{60mm} - B_0(m_{A^0}^2, m_{\nt_k}^2, m_{\nt_l}^2) \bigg]
\\[2mm]\non 
\delta Z_{33}^{H,\chp} &=& \frac{1}{(4\pi)^2}\, g^2 
\sum_{k,l=1}^2 \bigg[ \Big( (F^+_{kl3})^2 + (F^+_{lk3})^2 \Big) 
\bigg( \! \Big( m_{\chp_k}^2 \!+\! m_{\chp_l}^2 \!-\! m_{A^0}^2 
\Big) \dot{B}_0 - B_0 \bigg) 
\\
&&\hspace{40mm} - 4 m_{\chp_k} m_{\chp_l} F^+_{kl3} \, F^+_{lk3}
\, \dot{B}_0 \bigg] (m_{A^0}^2, m_{\chp_k}^2, m_{\chp_l}^2)
\\[2mm]\non 
\delta Z_{33}^{H,H} &=& -\frac{1}{(4\pi)^2}\, 
\Big(\frac{g_{\scriptscriptstyle Z}\,m_{\scriptscriptstyle 
Z}}{2}\Big)^2 \, \sum_{k=1}^2  \sum_{l=3}^4 \Big( A_{k,l-2} 
\Big)^2\, \dot{B}_0(m_{A^0}^2, m_{H_k^0}^2, m_{H_l^0}^2) 
\\[2mm]
&&\hspace{40mm}- \frac{1}{(4\pi)^2}\, 2\, 
\Big(\frac{g\,m_{\scriptscriptstyle W}}{2}\Big)^2\, 
\dot{B}_0(m_{A^0}^2, m_{H^+}^2, m_{W^+}^2) 
\\[2mm]\non 
\delta Z_{33}^{H,Z} &=& \frac{1}{(4\pi)^2}\, 
\frac{g_{\scriptscriptstyle Z}^2}{4} \sum_{k=1}^2 \Big( 
R_{1k}(\a\!-\!\b) \Big)^2 \bigg[ \dot{B}_0(m_{A^0}^2, m_{H_k^0}^2, 
m_{Z}^2) \Big( 2 m_{A^0}^2 \!+\! 2 m_{H_k^0}^2 \!-\! 
m_{\scriptscriptstyle Z}^2 \Big) 
\\
&&\hspace{60mm}+2 B_0(m_{A^0}^2, m_{H_k^0}^2, m_{Z}^2) \bigg]
\\[2mm]\non 
\delta Z_{33}^{H,W} &=& \frac{1}{(4\pi)^2}\,2\, \frac{g^2}{4} 
\bigg[ \dot{B}_0(m_{A^0}^2, m_{H^+}^2, m_{W}^2) \Big( 2 m_{A^0}^2 
\!+\! 2 m_{H^+}^2 \!-\! m_{\scriptscriptstyle W}^2 \Big) 
\\
&&\hspace{60mm}+2 B_0(m_{A^0}^2, m_{H^+}^2, m_{W}^2) \bigg]
\end{eqnarray}

\subsection{Sfermion part}
\begin{eqnarray}\non
\delta Z_{ii}^{\sf,\, \tilde\chi} &=& \vor \, \sum_{k=1}^4 \bigg[
\Big( (a^\sf_{ik})^2 + (b^\sf_{ik})^2 \Big) \cdot \Big(
(m_{\nt_k}^2 + m_f^2 - m_{\sf_i}^2)\, \dot{B}_0 - B_0 \Big)
\\[2mm]\non
&&\hspace{45mm} +4 m_{\nt_k} m_f \, a^\sf_{ik} b^\sf_{ik} \,
\dot{B}_0  \bigg] (m_{\sf_i}^2, m_{\nt_k}^2, m_f^2)
\\[2mm]\non
&&\hspace{-3mm}+\vor \, \sum_{k=1}^2 \bigg[ \Big( (k^\sf_{ik})^2 +
(l^\sf_{ik})^2 \Big) \cdot \Big( (m_{\chp_k}^2 + m_{f'}^2 -
m_{\sf_i}^2)\, \dot{B}_0 - B_0 \Big)
\\[2mm]
&&\hspace{45mm} +4 m_{\chp_k} m_{f'} \, k^\sf_{ik} l^\sf_{ik} \,
\dot{B}_0  \bigg] (m_{\sf_i}^2, m_{\chp_k}^2, m_{f'}^2)
\\[2mm]\non 
\delta Z_{ii}^{\sf,\, H} &=& -\vor \, \sum_{k=1}^4 \sum_{m=1}^2 
G^\sf_{mik}\, G^\sf_{imk}\, \dot{B}_0(m_{\sf_i}^2, m_{\sf_m}^2, 
m_{H_k^0}^2) 
\\[2mm]
&&-\vor \, \sum_{k=1}^2 \sum_{m=1}^2 G^{\sf'\sf}_{mik} \,
G^{\sf\sf'}_{imk}\, \dot{B}_0(m_{\sf_i}^2, m_{\sf'_m}^2,
m_{H_k^+}^2)
\\[2mm]\non 
\delta Z_{ii}^{\sf,\, \g} &=& \vor \, (e_0 e_f)^2 \bigg[ 2 B_0 + 
\Big( 4 m_{\sf_i}^2 \!-\! \l^2 \Big) \dot{B}_0 \bigg] 
(m_{\sf_i}^2, m_{\sf_i}^2, \l^2) 
\\[2mm]
\delta Z_{ii}^{\sf,\, Z} &=& \vor \, g_{\scriptscriptstyle Z}^2 
\sum_{m=1}^2 \Big(z^\sf_{im}\Big)^2 \bigg[ 2 B_0 + \Big( 2 
m_{\sf_i}^2 \!+\! 2 m_{\sf_m}^2 \!-\! m_{\scriptscriptstyle Z}^2 
\Big) \dot{B}_0 \bigg] (m_{\sf_i}^2, m_{\sf_m}^2, 
m_{\scriptscriptstyle Z}^2) 
\\[2mm]\non
\delta Z_{ii}^{\sf,\, W} &=& \vor \, \frac{g^2}{2} \sum_{m=1}^2
\Big( R^{\sf}_{i1} R^{\sf'}_{m1} \Big)^2 \bigg[ 2 B_0 + \Big( 2
m_{\sf_i}^2 \!+\! 2 m_{\sf'_m}^2 \!-\! m_{\scriptscriptstyle W}^2
\Big) \dot{B}_0 \bigg] (m_{\sf_i}^2, m_{\sf'_m}^2,
m_{\scriptscriptstyle W}^2)
\\
\end{eqnarray}

\clearpage
\section{Self--energies and counter terms}\label{appSE}
Here we give the explicit form of the self--energies needed for 
the computation of various counter terms for the one--loop width 
$A^0 \rightarrow \sf_1 \,\bar{\sf_2}$.
\subsection{\boldmath{$AZ$}--mixing}\label{appAZmixing}
The scalar--vector mixing self--energy, $\Pi_{AZ}(k^2)$, is 
defined by the two--point function \\
\begin{picture}(140,30)(0,0)
     \put(15,0){\mbox{\resizebox{13cm}{!}
     {\includegraphics{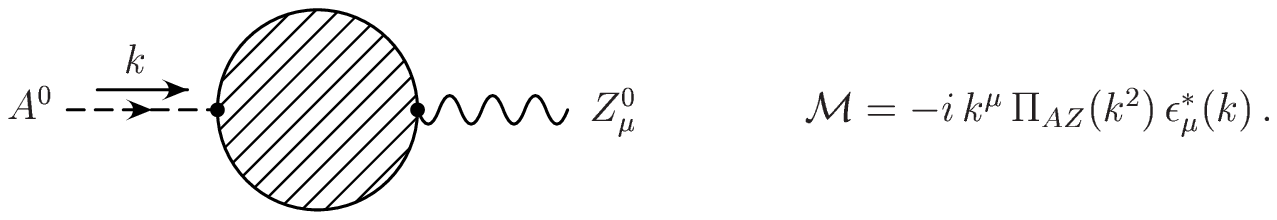}}}}
\end{picture}\\[4mm]
The single contributions from the particles $x$ are denoted by the 
superscript $x$ in $\Pi_{AZ}^x$. 
\begin{eqnarray}
\Pi_{AZ}^f &=& - \frac{i}{(4\pi)^2}\, m_{\scriptscriptstyle Z}
\sin 2\b \sum_{f} N_C^f\, I^{3L}_f \, h_f^2\, B_0(m_{A^0}^2,
m_f^2, m_f^2)
\\[2mm]
\Pi_{AZ}^{\nt} &=& \frac{i}{(4\pi)^2}\, 2\, g\, 
g_{\scriptscriptstyle Z} \sum_{k,l=1}^4 \!F^0_{kl3}\, O^{''L}_{lk} 
\Big[ m_{\nt_l}\, B_0 + (m_{\nt_l}\!-\!m_{\nt_k}) B_1 \Big] 
(m_{A^0}^2, m_{\nt_l}^2, m_{\nt_k}^2) 
\\[2mm]\non 
\Pi_{AZ}^{\chp} &=& \frac{i}{(4\pi)^2}\, 2\, g\, 
g_{\scriptscriptstyle Z} \sum_{k,l=1}^2 \bigg[ \Big( F^+_{kl3}\, 
O^{'L}_{lk} - F^+_{lk3}\, O^{'R}_{lk} \Big) m_{\chp_l} (B_0 + B_1) 
\\
&&\hspace{35mm}+ \Big( F^+_{kl3}\, O^{'R}_{lk} - F^+_{lk3}\,
O^{'L}_{lk} \Big) m_{\chp_k}\, B_1 \bigg] (m_{A^0}^2,
m_{\chp_l}^2, m_{\chp_k}^2)
\\[2mm] 
\Pi_{AZ}^\sf &=& -\vor\, 2\,g_{\scriptscriptstyle Z} \sum_{f} 
N_C^f\,z^\sf_{21}\, G^\sf_{123} \Big( B_0+2 B_1 \Big) (m_{A^0}^2, 
m_{\sf_1}^2, m_{\sf_2}^2) 
\\[2mm] 
\Pi_{AZ}^H &=& \frac{i}{(4\pi)^2}\, \frac{g_{\scriptscriptstyle 
Z}^2\, m_{\scriptscriptstyle Z}}{4} \sum_{k=1}^2 \sum_{l=3}^4 
A_{k,l-2}\, R_{k,l-2}(\b\!-\!\a) \Big( B_0+2 B_1 \Big) (m_{A^0}^2, 
m_{H_l^0}^2, m_{H_k^0}^2) 
\\[2mm] 
\Pi_{AZ}^Z &=& \frac{i}{(4\pi)^2}\, \frac{g_{\scriptscriptstyle 
Z}^2\, m_{\scriptscriptstyle Z}}{4} \sin(2\a\!-\!2\b) \sum_{k=1}^2 
(-1)^k \Big( B_0 - B_1 \Big) (m_{A^0}^2, m_{H_k^0}^2, 
m_{\scriptscriptstyle Z}^2) 
\end{eqnarray}

\clearpage 
\begin{figure}[th]
\begin{picture}(170,105)(0,0)
     \put(0,0){\mbox{\resizebox{16cm}{!}
     {\includegraphics{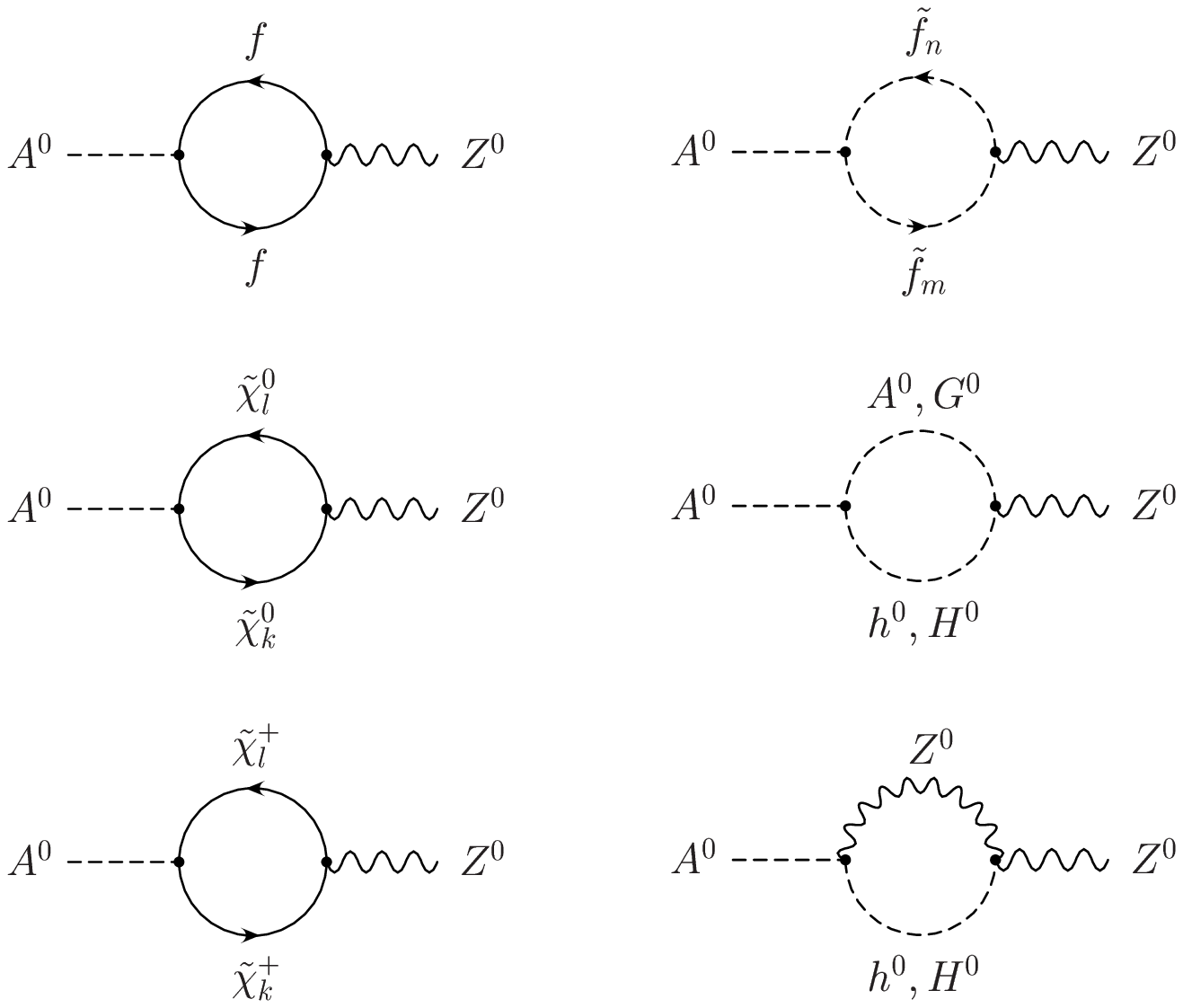}}}}
\end{picture}
\caption{$AZ$--mixing self--energies\label{AZmixing}}
\end{figure}

\subsection{\boldmath{$W^+$} self--energies}\label{appW-SE}
For the calculation of the mass counter term of a gauge boson $V$
$(V = W^\pm, Z^0)$, $\d m_V^2 = \Re\, \Pi^T_{VV} (m_V^2)$, we need 
the transverse part of the vector self--energy $\Pi^T_{VV} (k^2)$ 
from
\begin{eqnarray}\label{VVSE_def}
\begin{picture}(140,25)(0,0)
     \put(-5,0){\mbox{\resizebox{16cm}{!}
     {\includegraphics{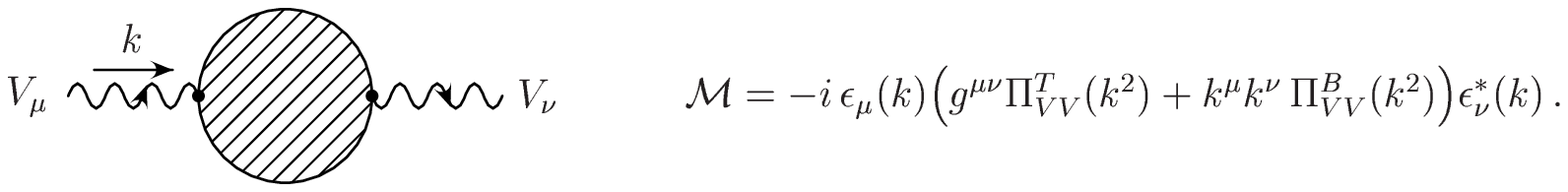}}}}
\end{picture}
\end{eqnarray}
\begin{eqnarray}\non
\left( \frac{\d m_{\scriptscriptstyle W}}{m_{\scriptscriptstyle 
W}} \right)^{ff} &=& -\frac{1}{(4 \pi)^2} \sum_{\rm gen.} N_C^f 
\Bigg[ h_{f_\uparrow}^2 \sin^2\!\b \, 
\frac{A_0(m_{f_\uparrow}^2)}{m_{f_\uparrow}^2} + 
h_{f_\downarrow}^2 \cos^2\!\b \, B_0( m_{\scriptscriptstyle W}^2, 
m_{f_\downarrow}^2, m_{f_\uparrow}^2 ) 
\\
&&\hspace{15mm} -~\frac{g^2}{m_{\scriptscriptstyle W}^2} B_{00}( 
m_{\scriptscriptstyle W}^2, m_{f_\downarrow}^2, m_{f_\uparrow}^2 ) 
+ \frac{g^2}{2} B_1( m_{\scriptscriptstyle W}^2, 
m_{f_\downarrow}^2, m_{f_\uparrow}^2 ) \Bigg] 
\\[2mm] 
\left( \frac{\d m_{\scriptscriptstyle W}}{m_{\scriptscriptstyle 
W}} \right)^{\sf\sf} &=& -\frac{1}{(4 \pi)^2} \, 
\frac{g^2}{m_{\scriptscriptstyle W}^2} \sum_{\rm gen.} N_C^f 
\sum_{m,n =1}^2 \Big( R^{\sf_{\uparrow}}_{m1} 
R^{\sf_{\downarrow}}_{n1} \Big)^2 B_{00}(m_{\scriptscriptstyle 
W}^2, m_{\sf_{\uparrow m}}^2, m_{\sf_{\downarrow n}}^2) 
\end{eqnarray}
Here $f_\uparrow$ and $f_\downarrow$ denote up-- and down--type 
(s)fermions of all three generations, respectively. 
\begin{eqnarray}
\left( \frac{\d m_{\scriptscriptstyle W}}{m_{\scriptscriptstyle 
W}} \right)^{\sf} &=& \vor \, \frac{g^2}{4 m_{\scriptscriptstyle 
W}^2} \sum_{\rm gen.} N_C^f \sum_{m=1}^2 \Big( R^\sf_{m1} \Big)^2 
A_0(m_{\sf_{m}}^2) 
\\[2mm] \non 
\left( \frac{\d m_{\scriptscriptstyle 
W}}{m_{\scriptscriptstyle W}} \right)^{\ti\chi} &=& \vor \, 
\frac{g^2}{m_{\scriptscriptstyle W}^2} \sum_{k=1}^2 \sum_{l=1}^4 
\bigg[ 2\, O^L_{lk}\, O^R_{lk}\, m_{\chp_k}\, m_{\nt_l}\, B_0 - 
\Big( (O^L_{lk})^2 + (O^R_{lk})^2 \Big)\, \times 
\\
&&\hspace{7mm} \Big( m_{\scriptscriptstyle W}^2\, B_1 + 
m_{\nt_l}^2\, B_0 + A_0(m_{\chp_k}^2) - 2 \, B_{00} \Big) \bigg] 
(m_{\scriptscriptstyle W}^2, m_{\nt_l}^2, m_{\chp_k}^2) 
\\[2mm]\non
\left( \frac{\d m_{\scriptscriptstyle W}}{m_{\scriptscriptstyle 
W}} \right)^{HH} &=& -\vor \, \frac{g^2}{2 m_{\scriptscriptstyle 
W}^2} \Bigg[\, \sum_{k,l=1}^2 \Big( R_{lk}(\a\!-\!\b) \Big)^2\, 
B_{00}(m_{\scriptscriptstyle W}^2, m_{H_l^+}^2, m_{H_k^0}^2) 
\\
&&\hspace{25mm} +~B_{00}(m_{\scriptscriptstyle W}^2, m_{H^+}^2,
m_{A^0}^2) + B_{00}(m_{\scriptscriptstyle W}^2, m_{G^+}^2,
m_{G^0}^2) \Bigg]
\\[2mm] 
\left( \frac{\d m_{\scriptscriptstyle W}}{m_{\scriptscriptstyle 
W}} \right)^{H} &=& \vor \, \frac{g^2}{8 m_{\scriptscriptstyle 
W}^2} \left(\ \sum_{k=1}^4 A_0(m_{H_k^0}^2) + 2 \sum_{k=1}^2 
A_0(m_{H_k^+}^2) \right) 
\\[2mm]\non 
\left( \frac{\d m_{\scriptscriptstyle 
W}}{m_{\scriptscriptstyle W}} \right)^{VS} &=& \vor \, 
\frac{g^2}{2} \Bigg[ \ \sum_{k=1}^2 \Big( R_{2k}(\a\!-\!\b) 
\Big)^2 B_0(m_{\scriptscriptstyle W}^2, m_{H_k^0}^2, 
m_{\scriptscriptstyle W}^2) 
\\
&&\hspace{15mm} +~s_{\scriptscriptstyle W}^2 \,
B_0(m_{\scriptscriptstyle W}^2, m_{\scriptscriptstyle W}^2, \l^2)
+ s_{\scriptscriptstyle W}^2 t_{\scriptscriptstyle W}^2 \,
B_0(m_{\scriptscriptstyle W}^2, m_{\scriptscriptstyle W}^2,
m_{\scriptscriptstyle Z}^2) \Bigg]
\\[2mm]\non
\left( \frac{\d m_{\scriptscriptstyle W}}{m_{\scriptscriptstyle 
W}} \right)^{VV+V+\rm ghost} \hspace{-2mm}&=& -\vor \, 
\frac{g^2}{2 m_{\scriptscriptstyle W}^2} \Bigg[ 
s_{\scriptscriptstyle W}^2 \bigg(\! 8 B_{00} + 7 
m_{\scriptscriptstyle W}^2 B_0 + 2 m_{\scriptscriptstyle W}^2 B_1 
\! \bigg) (m_{\scriptscriptstyle W}^2, m_{\scriptscriptstyle W}^2, 
\l^2) 
\\ \non
&&\hspace{23mm} +c_{\scriptscriptstyle W}^2 \bigg(\! 8 B_{00} + 7
m_{\scriptscriptstyle W}^2 B_0 + 2 m_{\scriptscriptstyle W}^2 B_1
\! \bigg) (m_{\scriptscriptstyle W}^2, m_{\scriptscriptstyle W}^2,
m_{\scriptscriptstyle Z}^2)
\\
&&\hspace{35mm} - s_{\scriptscriptstyle W}^2 A_0(\l^2) -
c_{\scriptscriptstyle W}^2 A_0(m_{\scriptscriptstyle Z}^2) - 3
A_0(m_{\scriptscriptstyle W}^2) \Bigg]
\end{eqnarray}

\begin{figure}[th]
\begin{picture}(170,200)(0,0)
     \put(0,0){\mbox{\resizebox{16cm}{!}{\includegraphics{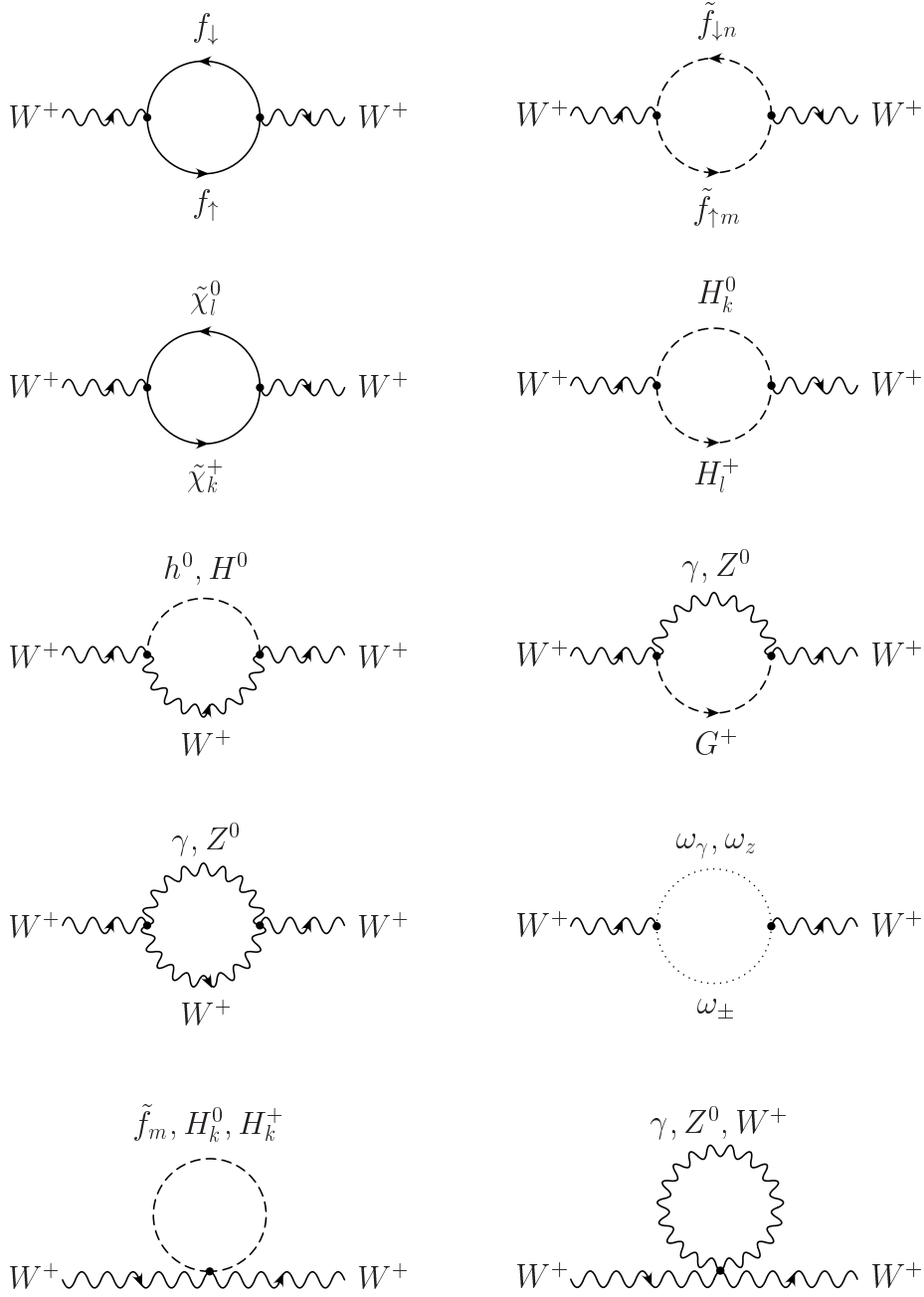}}}}
\end{picture}
\caption{$W^+$ self--energies\label{W-SE}}
\end{figure}

\clearpage
\subsection{\boldmath{$Z^0$} self--energies}\label{appZ-SE}
Accordingly to eq.~(\ref{VVSE_def}) the mass counter term 
contributions to the gauge boson $Z^0$ are:
\begin{eqnarray}\non
\left( \frac{\d m_{\scriptscriptstyle Z}}{m_{\scriptscriptstyle
Z}} \right)^{ff} &=& \frac{1}{(4 \pi)^2} \,
\frac{g_{\scriptscriptstyle Z}^2}{m_{\scriptscriptstyle Z}^2} 
\sum_{f} N_C^f \bigg[ 2\, C^f_L\, C^f_R \, m_f^2 \, B_0 - \Big( 
(C^f_L)^2 + (C^f_R)^2 \Big)\, \times 
\\
&&\hspace{12mm} \bigg( A_0(m_f^2) + m_f^2 \, B_0 - 2 B_{00} +
m_{\scriptscriptstyle Z}^2 \, B_1 \bigg) \bigg] \Big(
m_{\scriptscriptstyle Z}^2, m_f^2, m_f^2 \Big)
\\[2mm]
\left( \frac{\d m_{\scriptscriptstyle Z}}{m_{\scriptscriptstyle 
Z}} \right)^{\sf\sf} &=& -\vor \, \frac{2\, g_{\scriptscriptstyle 
Z}^2}{m_{\scriptscriptstyle Z}^2} \sum_{f} N_C^f \sum_{m,n =1}^2 
\Big( z^\sf_{mn} \Big)^2 \, B_{00}(m_{\scriptscriptstyle Z}^2, 
m_{\sf_m}^2, m_{\sf_n}^2) 
\\[2mm]
\left( \frac{\d m_{\scriptscriptstyle Z}}{m_{\scriptscriptstyle 
Z}} \right)^{\sf} &=& \vor \, \frac{g_{\scriptscriptstyle 
Z}^2}{m_{\scriptscriptstyle Z}^2} \sum_{f} N_C^f \sum_{m=1}^2 
\bigg( \Big(C^f_L\,R^\sf_{m1}\Big)^2 + \Big(C^f_R 
\,R^\sf_{m2}\Big)^2 \bigg) A_0(m_{\sf_{m}}^2) 
\\[2mm]\non
\left( \frac{\d m_{\scriptscriptstyle Z}}{m_{\scriptscriptstyle 
Z}} \right)^{\nt} &=& -\vor \, \frac{g_{\scriptscriptstyle 
Z}^2}{m_{\scriptscriptstyle Z}^2} \sum_{k,l=1}^4 \Big( O^{''}_{kl} 
\Big)^2 \bigg[ (m_{\nt_k} \!+\! m_{\nt_l}) m_{\nt_l}\, B_0 + 
m_{\scriptscriptstyle Z}^2\, B_1 
\\
&&\hspace{40mm} +A_0(m_{\nt_k}^2) -~2\, B_{00} \bigg]
(m_{\scriptscriptstyle Z}^2, m_{\nt_k}^2, m_{\nt_l}^2)
\\[2mm]\non
\left( \frac{\d m_{\scriptscriptstyle Z}}{m_{\scriptscriptstyle 
Z}} \right)^{\chp} &=& \vor \, \frac{g_{\scriptscriptstyle 
Z}^2}{m_{\scriptscriptstyle Z}^2} \sum_{k,l=1}^2 \bigg[ 2\, 
O^{'L}_{kl}\, O^{'R}_{kl}\, m_{\chp_k}\, m_{\chp_l}\, B_0 - \Big( 
(O^{'L}_{kl})^2 + (O^{'R}_{kl})^2 \Big)\, \times 
\\
&&\hspace{4mm} \Big( m_{\scriptscriptstyle Z}^2\, B_1 +
m_{\chp_k}^2\, B_0 + A_0(m_{\chp_k}^2) - 2 \, B_{00} \Big) \bigg]
(m_{\scriptscriptstyle Z}^2, m_{\chp_k}^2, m_{\chp_l}^2)
\\[2mm]\non
\left( \frac{\d m_{\scriptscriptstyle Z}}{m_{\scriptscriptstyle 
Z}} \right)^{HH} &=& -\vor \, \frac{g_{\scriptscriptstyle Z}^2}{2 
m_{\scriptscriptstyle Z}^2} \Bigg[ \sum_{k=1}^2 \sum_{l=3}^4 \Big( 
R_{k,l-2}(\b\!-\!\a) \Big)^2 \, B_{00}(m_{\scriptscriptstyle Z}^2, 
m_{H_k^0}^2, m_{H_l^0}^2) 
\\
&&\hspace{37mm} + \cos^2(2\tw) \sum_{k=1}^2
B_{00}(m_{\scriptscriptstyle Z}^2, m_{H_k^+}^2, m_{H_k^+}^2)
\Bigg]
\\[2mm]
\left( \frac{\d m_{\scriptscriptstyle Z}}{m_{\scriptscriptstyle 
Z}} \right)^{H} &=& \vor \, \frac{g_{\scriptscriptstyle Z}^2}{8 
m_{\scriptscriptstyle Z}^2} \Bigg[ \sum_{k=1}^4 A_0(m_{H_k^0}^2) + 
2 \cos^2(2 \tw) \sum_{k=1}^2 A_0(m_{H_k^+}^2) \Bigg] 
\\[2mm]\non
\left( \frac{\d m_{\scriptscriptstyle Z}}{m_{\scriptscriptstyle 
Z}} \right)^{VS} &=& \vor \Bigg( \frac{g_{\scriptscriptstyle 
Z}^2}{2} \sin^2(\a\!-\!\b)\, B_0(m_{\scriptscriptstyle Z}^2, 
m_{h^0}^2, m_{\scriptscriptstyle Z}^2) + 
\frac{g_{\scriptscriptstyle Z}^2}{2} \cos^2(\a\!-\!\b) \, \times 
\\
&&\hspace{21mm} B_0(m_{\scriptscriptstyle Z}^2, m_{H^0}^2,
m_{\scriptscriptstyle Z}^2) + g^2\, s_{\scriptscriptstyle W}^4 \,
B_0(m_{\scriptscriptstyle Z}^2, m_{W}^2, m_{G^+}^2) \Bigg)
\\[2mm]\non
\left( \frac{\d m_{\scriptscriptstyle Z}}{m_{\scriptscriptstyle 
Z}} \right)^{WW+W+ \rm ghost} &=& -\vor \, \frac{g^2 
c_{\scriptscriptstyle W}^2}{m_{\scriptscriptstyle Z}^2} \bigg[ 4 
B_{00} + m_{\scriptscriptstyle W}^2 B_0 + \frac{5}{2} 
m_{\scriptscriptstyle Z}^2 B_0 + m_{\scriptscriptstyle Z}^2 B_1 
\\
&&\hspace{40mm} - 2 A_0(m_{\scriptscriptstyle W}^2) \!
\bigg](m_{\scriptscriptstyle Z}^2, m_{\scriptscriptstyle W}^2,
m_{\scriptscriptstyle W}^2)
\end{eqnarray}

\begin{figure}[th]
\begin{picture}(170,200)(0,0)
     \put(0,0){\mbox{\resizebox{16cm}{!}{\includegraphics{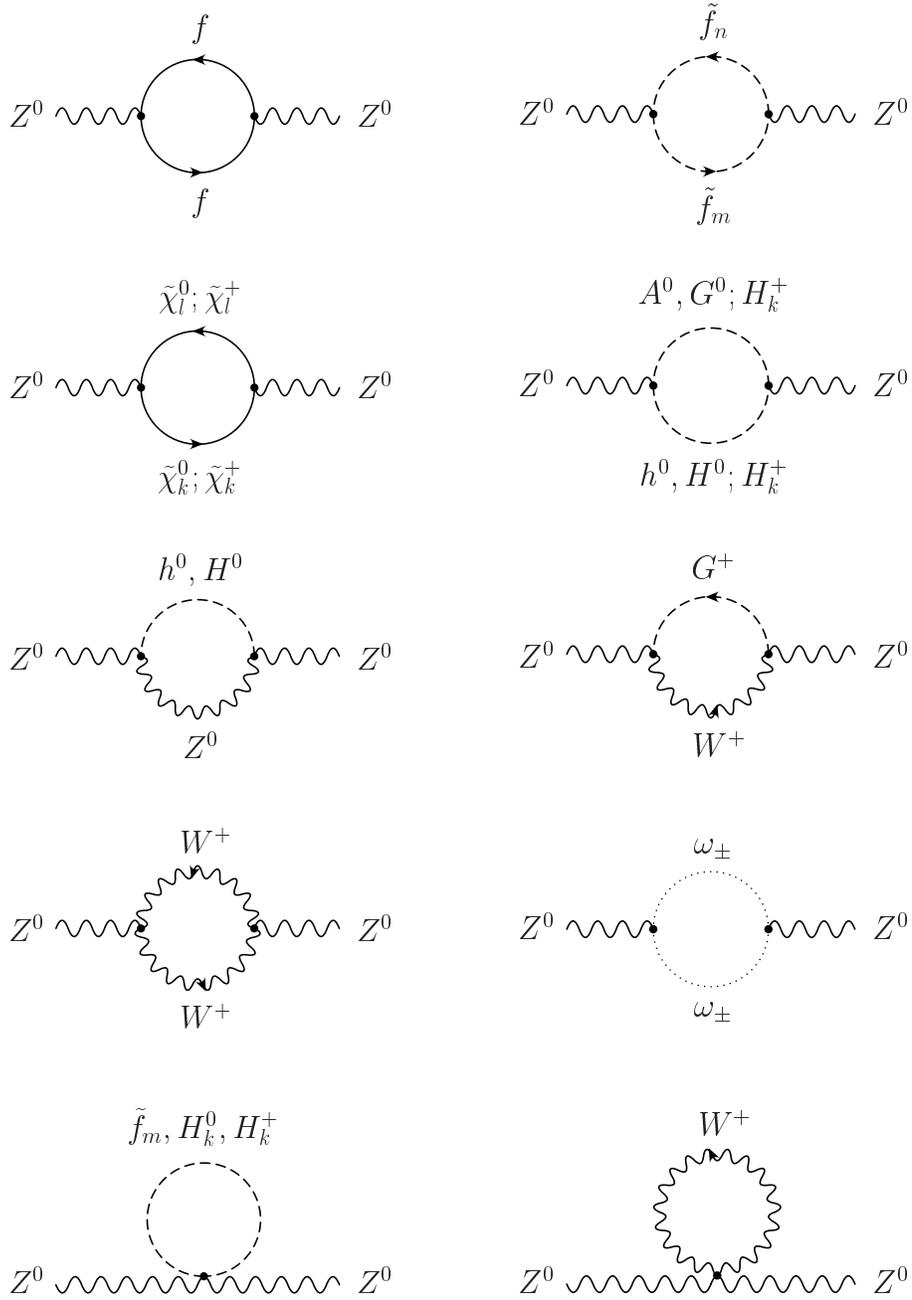}}}}
\end{picture}
\caption{$Z^0$ self--energies\label{Z-SE}}
\end{figure}

\clearpage
\subsection{Fermion self--energies}\label{appfermion-SE}
In our notation, the fermion self--energy is defined by 
\\ 
\begin{picture}(140,25)(0,0)
     \put(15,0){\mbox{\resizebox{11.5cm}{!}
     {\includegraphics{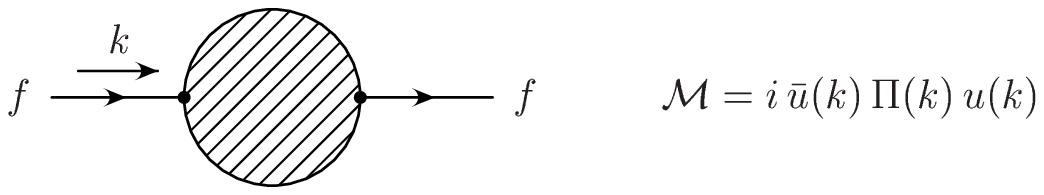}}}}
\end{picture}\\
with
\begin{eqnarray}\label{Piferm}
\Pi (k) & = & \not\!k \, P_L \, \Pi^L (k) + \not\!k \, P_R \, 
\Pi^R (k) + \Pi^{SL} (k) P_L + \Pi^{SR} (k) P_R \,. 
\end{eqnarray}
Thus the counter term for quarks and leptons is given by
\begin{eqnarray}
\d m_f & = & \frac{1}{2}\, \Re \left[ m_f \left( \Pi^L (m_f) + 
\Pi^R (m_f) \right) + \Pi^{SL}(m_f) + \Pi^{SR}(m_f) \right] \,. 
\end{eqnarray}
Note that for quarks and leptons (contrary to charginos), the 
left-- and right--handed scalar parts of $\Pi (k)$ are equal,
$\Pi^{SL}(k) = \Pi^{SR}(k)$. The single contributions to $\d m_f$ 
are as follows: 

\begin{eqnarray}\non
\bigg( \frac{\d m_f}{m_f} \bigg)^{f H_k^0} &=& \vor \Bigg[
\sum_{k=1}^2 (s^f_k)^2 \Big( B_0 - B_1 \Big) + \sum_{k=3}^4
(s^f_k)^2 \, \Big( B_0 + B_1 \Big) \Bigg] (m_f^2, m_f^2,
m_{H_k^0}^2)
\\[-3mm]
\\ \non
\bigg( \frac{\d m_f}{m_f} \bigg)^{f' H_k^+} &=& -\vor \sum_{k=1}^2
\Bigg[ \frac{1}{2} \bigg( (y^f_{k})^2 + (y^{f'}_{k})^2 \bigg) B_1 
- \frac{m_{f'}}{m_f}\, y^f_{k}\, y^{f'}_{k} \,B_0 \Bigg] (m_f^2, 
m_{f'}^2, m_{H_k^+}^2) 
\\
\\[2mm]\non
\bigg( \frac{\d m_f}{m_f} \bigg)^{\sf\,\nt} &=& -\vor \sum_{m=1}^2 
\sum_{k=1}^4 \Bigg[ \frac{1}{2} \bigg( (a^\sf_{mk})^2 + 
(b^\sf_{mk})^2 \bigg) B_1 
\\
&&\hspace{40mm} - \frac{m_{\nt_k}}{m_f}\, a^\sf_{mk} b^\sf_{mk}
\,B_0 \Bigg] (m_f^2, m_{\nt_k}^2, m_{\sf_m}^2)
\\[2mm]\non
\bigg( \frac{\d m_f}{m_f} \bigg)^{\sf'\chp} &=& -\vor \sum_{m=1}^2 
\sum_{k=1}^2 \Bigg[ \frac{1}{2} \bigg( (k^{\sf'}_{mk})^2 + 
(l^{\sf'}_{mk})^2 \bigg) B_1 
\\
&&\hspace{40mm} - \frac{m_{\chp_k}}{m_f}\, k^{\sf'}_{mk}\,
l^{\sf'}_{mk} \,B_0 \Bigg] (m_f^2, m_{\chp_k}^2, m_{\sf'_m}^2)
\\[2mm]\non
\bigg( \frac{\d m_f}{m_f} \bigg)^{f \g} &=& -\vor \, 2 (e_0 e_f)^2 
\Big( B_0 - B_1 \Big) (m_f^2, \l^2, m_{f}^2) 
\\[2mm]
\bigg( \frac{\d m_f}{m_f} \bigg)^{f Z^0} &=& -\vor \, 
g_{\scriptscriptstyle Z}^2 \bigg[ \Big( (C^f_L)^2 + (C^f_R)^2 
\Big) B_1 + 4 C^f_L C^f_R \, B_0 \bigg] (m_f^2, m_f^2, 
m_{\scriptscriptstyle Z}^2) 
\\[2mm]
\bigg( \frac{\d m_f}{m_f} \bigg)^{f' W^+} &=& -\vor \, 
\frac{g^2}{2} \, B_1 (m_f^2, m_{f'}^2, m_{\scriptscriptstyle W}^2) 
\end{eqnarray}

\clearpage 
\begin{figure}[th]
\begin{picture}(170,105)(0,0)
     \put(0,0){\mbox{\resizebox{16cm}{!}
     {\includegraphics{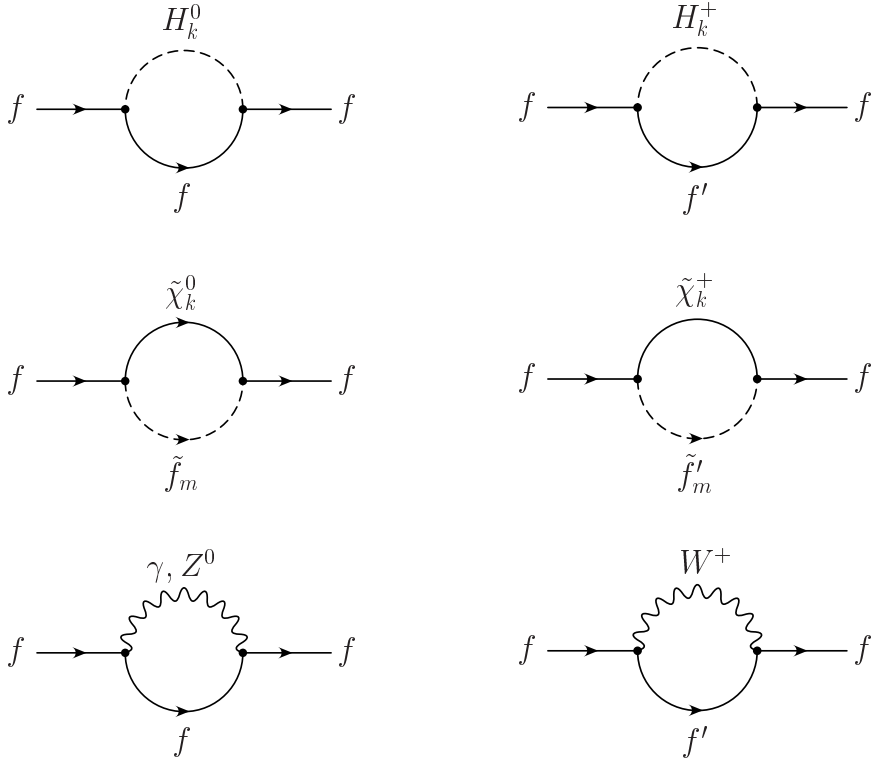}}}}
\end{picture}
\caption{Fermion self--energies\label{fermion-SE}}
\end{figure}

\subsection{Chargino self--energies}\label{appchargino-SE}
Since the higgsino mass parameter $\mu$ is fixed in the chargino 
sector, the counter term $\d \mu$ reads \cite{0104109, Willi} 
\begin{eqnarray}
\d \mu &=& \d X_{22} ~=~ \frac{1}{2}\, \sum_{k,l=1}^2 U_{k2} 
V_{l2} \Big( \Pi^L_{lk}\, m_{\chp_k} + \Pi^R_{kl}\, m_{\chp_l} + 
\Pi^{SL}_{kl} + \Pi^{SR}_{lk} \Big) 
\end{eqnarray}
with the chargino self--energies $\Pi_{kl} = 
\Pi_{kl}(m_{\chp_l}^2)$. $U$ and $V$ are two real $2\!\times\!2$ 
matrices which diagonalize the chargino mass matrix, 
\begin{eqnarray}\non
U\, X\, V^T &=& M_D ~=~ \bmat m_{\tilde\chi^+_1} & 0 \\ 0 & 
m_{\tilde\chi^+_2} \emat \,.
\end{eqnarray}
The single left-- and right--handed parts of $\Pi_{kl}$ can be 
found by comparing the coefficients accordingly to 
eq.~(\ref{Piferm}).\\

\noindent fermion--sfermion contribution:
\begin{eqnarray}\non
\Pi_{ij}^{L, f}(k^2) &=& -\vor \, \sum_f N_C^f \sum_{m=1}^2 \bigg[
k_{mi}^{\sf_\downarrow} k_{mj}^{\sf_\downarrow}\, B_1\!\Big(k^2,
m_{f_\uparrow}^2, m_{\sf_{\downarrow m}}^2\Big) +
l_{mi}^{\sf_\uparrow} l_{mj}^{\sf_\uparrow} \,B_1\!\Big(k^2,
m_{f_\downarrow}^2, m_{\sf_{\uparrow m}}^2\Big) \bigg]
\\[2mm] \non
\Pi_{ij}^{R, f}(k^2) &=& -\vor \, \sum_f N_C^f \sum_{m=1}^2 \bigg[
l_{mi}^{\sf_\downarrow} l_{mj}^{\sf_\downarrow}\, B_1\!\Big(k^2,
m_{f_\uparrow}^2, m_{\sf_{\downarrow m}}^2\Big) +
k_{mi}^{\sf_\uparrow} k_{mj}^{\sf_\uparrow} \,B_1\!\Big(k^2,
m_{f_\downarrow}^2, m_{\sf_{\uparrow m}}^2\Big) \bigg]
\\[2mm] \non
\Pi_{ij}^{SL, f}(k^2) &=& \hspace{3.25mm} \vor \, \sum_f N_C^f
\sum_{m=1}^2 \bigg[ m_{f_{\uparrow}} l_{mi}^{\sf_\downarrow}
k_{mj}^{\sf_\downarrow}\, B_0\Big(k^2, m_{f_\uparrow}^2,
m_{\sf_{\downarrow m}}^2\Big) + m_{f_{\downarrow}}
k_{mi}^{\sf_\uparrow} l_{mj}^{\sf_\uparrow} \, \times
\\[-2mm] \non
&& \hspace{96.5mm} B_0\Big(k^2, m_{f_\downarrow}^2,
m_{\sf_{\uparrow m}}^2\Big) \bigg]
\\[2mm] \non
\Pi_{ij}^{SR, f}(k^2) &=& \hspace{3.25mm} \vor \, \sum_f N_C^f
\sum_{m=1}^2 \bigg[ m_{f_{\uparrow}} k_{mi}^{\sf_\downarrow}
l_{mj}^{\sf_\downarrow}\, B_0\Big(k^2, m_{f_\uparrow}^2,
m_{\sf_{\downarrow m}}^2\Big) + m_{f_{\downarrow}}
l_{mi}^{\sf_\uparrow} k_{mj}^{\sf_\uparrow} \, \times
\\[-2mm] \non
&& \hspace{96.5mm} B_0\Big(k^2, m_{f_\downarrow}^2,
m_{\sf_{\uparrow m}}^2\Big) \bigg]
\\[2mm]
\end{eqnarray}
Higgs/gaugino contribution:
\begin{eqnarray}\non
\Pi^{H_l^0}_{ij}(k) &=& -\vor\, g^2 \sum_{k=1}^2 \left[ \kslash
\sum_{l=1}^4 \Big( F^+_{ikl} F^+_{jkl} \, P_L + F^+_{kil}
F^+_{kjl} \, P_R \Big) B_1 \right.
\\ \non
&& \hspace{28mm} - m_{\chp_k} \sum_{l=1}^2 \Big( F^+_{kil}
F^+_{jkl} \, P_L + F^+_{ikl} F^+_{kjl} \, P_R \Big) B_0
\\ \non
&& \hspace{28mm} \left. + m_{\chp_k} \sum_{l=3}^4 \Big( F^+_{kil}
F^+_{jkl} \, P_L + F^+_{ikl} F^+_{kjl} \, P_R \Big) B_0\right] \!
\Big(k^2, m_{\chp_k}^2, m_{H_l^0}^2\Big)
\\
\end{eqnarray}
\begin{eqnarray}\non
\Pi^{H_l^+}_{ij}(k) &=& -\vor\, g^2 \sum_{k=1}^4 \sum_{l=1}^2
\bigg[ \!\kslash \Big( F^R_{ikl} F^R_{jkl} \, P_L + F^L_{ikl}
F^L_{jkl} \, P_R \Big) B_1
\\ \non
&& \hspace{34mm} - m_{\nt_k} \Big( F^L_{ikl} F^R_{jkl} \, P_L +
F^R_{ikl} F^L_{jkl} \, P_R \Big) B_0\bigg](k^2, m_{\nt_k}^2,
m_{H_l^+}^2\Big)
\\
\end{eqnarray}
\begin{eqnarray}
\Pi^{\g}_{ij}(k) &=& -\vor\, 2 e^2 \d_{ij} \Big[ \!\kslash B_1 + 2
m_{\chp_j} B_0 \, \Big](k^2, m_{\chp_j}^2, \l^2)
\end{eqnarray}
\begin{eqnarray}\non
\Pi^{Z^0}_{ij}(k) &=& -\vor\, 2 \,g_{\scriptscriptstyle Z}^2 
\sum_{k=1}^2 \bigg[ \!\kslash \Big( O^{'L}_{ik} O^{'L}_{kj} \, P_L 
+ O^{'R}_{ik} O^{'R}_{kj} \, P_R \Big) B_1 
\\ \non
&& \hspace{31mm} + 2 m_{\chp_k} \Big( O^{'R}_{ik} O^{'L}_{kj} \,
P_L + O^{'L}_{ik} O^{'R}_{kj} \, P_R \Big) B_0\bigg](k^2,
m_{\chp_k}^2, m_{\scriptscriptstyle Z}^2\Big)
\\
\end{eqnarray}
\begin{eqnarray}\non
\Pi^{W^+}_{ij}(k) &=& -\vor\, 2 \,g^2 \sum_{k=1}^4 \bigg[
\!\kslash \Big( O^{L}_{ki} O^{L}_{kj} \, P_L + O^{R}_{ki}
O^{R}_{kj} \, P_R \Big) B_1
\\ \non
&& \hspace{30mm} + 2 m_{\nt_k} \Big( O^{R}_{ki} O^{L}_{kj} \, P_L
+ O^{L}_{ki} O^{R}_{kj} \, P_R \Big) B_0\bigg](k^2, m_{\nt_k}^2,
m_{\scriptscriptstyle W}^2\Big)
\\
\end{eqnarray}

\clearpage 
\begin{figure}[th]
\begin{picture}(170,70)(0,0)
     \put(0,0){\mbox{\resizebox{16cm}{!}
     {\includegraphics{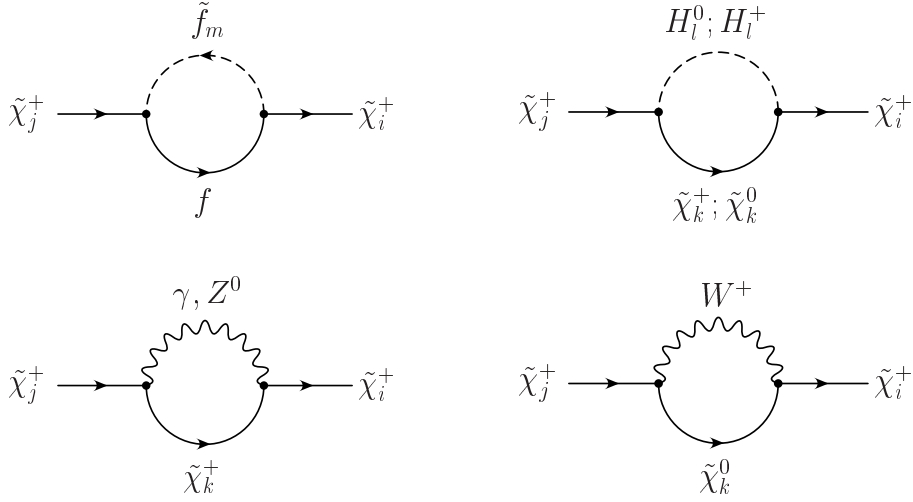}}}}
\end{picture}
\caption{Chargino self--energies\label{chargino-SE}}
\end{figure}

\subsection{Sfermion self--energies}\label{appsfermion-SE}
For the fixing of the sfermion mixing angle $\theta_\sf$ we need 
the off--diagonal elements of the sfermion self--energies, 
$\Pi_{ij}^{\sf} = \Pi_{ij}^{\sf} (m_{\sf_j}^2)$. In the following, 
$Y_{L/R}^f$ denotes the weak hypercharge, $Y_{L/R}^f \,=\, 2 
(I_f^{3L/R} - e_f)$. The short forms for various products of 
sfermion rotation matrices can be found in \ref{appVertex}. 
Additionally, we use the abbreviation $R^{\sf\ 
\!\hat{\!\!\sf}_{\!D}}_{ijkl} \,=\, R^\sf_{i1} R^\sf_{j1} R^{\ 
\!\hat{\!\!\sf}}_{k2} R^{\ \!\hat{\!\!\sf}}_{l2}$.
\begin{eqnarray}\non
\Pi_{ij}^{\sf,\, \tilde\chi} &=& -\vor \, \sum_{k=1}^4 \bigg[
\Big( a^\sf_{ik} a^\sf_{jk} + b^\sf_{ik} b^\sf_{jk} \Big) \cdot
\Big( A_0(m_{\nt_k}^2) + A_0(m_f^2) + (m_{\nt_k}^2 + m_f^2 -
m_{\sf_j}^2) \, B_0 \Big)
\\[2mm]\non
&&\hspace{20mm} + \Big( a^\sf_{ik} b^\sf_{jk} + b^\sf_{ik}
a^\sf_{jk} \Big) \cdot 2 m_{\nt_k} m_f B_0  \bigg] (m_{\sf_j}^2,
m_{\nt_k}^2, m_f^2)
\\[2mm]\non
&&-\vor \, \sum_{k=1}^2 \bigg[ \Big( k^\sf_{ik} k^\sf_{jk} +
l^\sf_{ik} l^\sf_{jk} \Big) \cdot \Big( A_0(m_{\chp_k}^2) +
A_0(m_{f'}^2) + (m_{\chp_k}^2 + m_{f'}^2 - m_{\sf_j}^2) \, B_0
\Big)
\\[2mm]
&&\hspace{20mm} + \Big( k^\sf_{ik} l^\sf_{jk} + l^\sf_{ik}
k^\sf_{jk} \Big) \cdot 2 m_{\chp_k} m_{f'} B_0  \bigg]
(m_{\sf_j}^2, m_{\chp_k}^2, m_{f'}^2)
\end{eqnarray}
\begin{eqnarray}\non
\Pi_{ij}^{\sf,\, H\sf} &=& \vor \, \sum_{k=1}^4 \sum_{m=1}^2
G^\sf_{mik}\, G^\sf_{jmk}\, B_0(m_{\sf_j}^2, m_{\sf_m}^2,
m_{H_k^0}^2)
\\[2mm]
&&\hspace{-3mm}+\vor \, \sum_{k=1}^2 \sum_{m=1}^2
G^{\sf'\sf}_{mik} \, G^{\sf\sf'}_{jmk}\, B_0(m_{\sf_j}^2,
m_{\sf'_m}^2, m_{H_k^+}^2)
\\[2mm]\non
\Pi_{ij}^{\sf,\, \g\sf} &=& -\vor \, (e_0 e_f)^2 \d_{ij} \bigg[ 2
A_0(\l^2) - A_0(m_{\sf_i}^2) + \Big( 4 m_{\sf_i}^2 \!-\! \l^2
\Big) B_0 (m_{\sf_i}^2, m_{\sf_i}^2, \l^2) \bigg]
\\
\\[2mm]\non
\Pi_{ij}^{\sf,\, Z\sf} &=& -\vor \, g_{\scriptscriptstyle Z}^2 
\sum_{m=1}^2 z^\sf_{mi} z^\sf_{jm} \bigg[ 2 
A_0(m_{\scriptscriptstyle Z}^2) -A_0(m_{\sf_m}^2) + \Big( 2 
m_{\sf_j}^2 \!+\! 2 m_{\sf_m}^2 \!-\! m_{\scriptscriptstyle Z}^2 
\Big) \, \times 
\\
&&\hspace{87mm}B_0 (m_{\sf_j}^2, m_{\sf_m}^2,
m_{\scriptscriptstyle Z}^2) \bigg]
\\[2mm]\non
\Pi_{ij}^{\sf,\, W\sf'} &=& -\vor \, \frac{g^2}{2} R^{\sf}_{i1}
R^{\sf}_{j1} \!\sum_{m=1}^2\! \Big( R^{\sf'}_{m1} \Big)^2 \bigg[ 2
A_0(m_{\scriptscriptstyle W}^2) -A_0(m_{\sf'_m}^2) + \Big( 2
m_{\sf_j}^2 \!+\! 2 m_{\sf'_m}^2 \!-\! m_{\scriptscriptstyle W}^2
\Big) \, \times
\\
&&\hspace{87mm}B_0 (m_{\sf_j}^2, m_{\sf'_m}^2,
m_{\scriptscriptstyle W}^2) \bigg]
\end{eqnarray}
\begin{eqnarray}\non
\Pi_{ij}^{\sf,\, \sf} &=&  \frac{1}{(4\pi)^2}\,h_f^2 \sum_{m=1}^2
\bigg[ N_C^f \Big( R^\sf_{jmmi} \!+\! R^\sf_{mijm} \Big) +
R^\sf_{jimm} + R^\sf_{mmji} \bigg] A_0(m_{\sf_m}^2)
\\[1mm] \non
&&\hspace{-3mm}+\frac{1}{(4\pi)^2}\, g_{\scriptscriptstyle Z}^2 
\sum_{m=1}^2 \Bigg\{ \bigg[ \Big( \frac{1}{4} - (2 I_f^{3L} \!-\! 
e_f) e_f s_{\scriptscriptstyle W}^2 \Big) R^{\sf_L}_{jimm} + e_f^2 
s_{\scriptscriptstyle W}^2 R^{\sf_R}_{jimm} \bigg] (N_C^f + 1) 
\\ \non
&&\hspace{25mm} + (I_f^{3L} \!-\! e_f) e_f s_{\scriptscriptstyle
W}^2 \bigg[ N_C^f \Big( R^{\sf}_{jimm} \!+\! R^{\sf}_{mmji} \Big)
\!+\! R^{\sf}_{jmmi} \!+\! R^{\sf}_{mijm} \bigg] \! \Bigg\}
A_0(m_{\sf_m}^2)
\\
\\\non
\Pi_{ij}^{\sf,\, \sf'} &=&  \frac{1}{(4\pi)^2} \sum_{m=1}^2 \Big( 
h_f^2 \, R^{\sf'\!\sf_D}_{mmji} + h_{f'}^2\, R^{\sf\sf'_D}_{jimm} 
\Big) A_0(m_{\sf'_m}^2) 
\\[1mm] \non
&&\hspace{-3mm}+\frac{1}{(4\pi)^2}\, \frac{g^2}{4} \sum_{m=1}^2 
\Bigg\{ N_C^f \bigg[ \Big( t_{\scriptscriptstyle W}^2 Y^f_L 
Y^{f'}_L \!-\! 1 \Big) R^{\sf\sf'_L}_{jimm} + 
t_{\scriptscriptstyle W}^2 Y^f_R Y^{f'}_R \, R^{\sf\sf'_R}_{jimm} 
\\[1mm]
&&\hspace{28mm}- Y^f_L Y^{f'}_R R^{\sf\sf'_D}_{jimm} - Y^{f'}_L 
Y^f_R R^{\sf'\!\sf_D}_{mmji} \bigg] +2 R^{\sf\sf'_L}_{jimm} 
\Bigg\} A_0(m_{\sf'_m}^2) 
\\[2mm]\non
\Pi_{ij}^{\sf,\, \ \!\hat{\!\!\sf}} &=&  \frac{1}{(4\pi)^2} 
N_C^{\hat f} \sum_{m=1}^2 \bigg[ h_f h_{\hat f} \Big( R^{\sf\ 
\!\hat{\!\!\sf}_{\!F}}_{ijmm} + R^{\sf\ 
\!\hat{\!\!\sf}_{\!F}}_{jimm} \Big) + \frac{g^2}{4} R^{\sf\ 
\!\hat{\!\!\sf}_{\!L}}_{jimm} \bigg] A_0(m_{\ 
\!\hat{\!\!\sf}_{\!m}}^2) 
\\[1mm] \non
&&\hspace{-3mm}+\frac{1}{(4\pi)^2}\, N_C^{\hat f}\, \frac{g'^2}{4}
\sum_{m=1}^2 \bigg[ Y^f_L Y^{\hat{f}}_L R^{\sf\
\!\hat{\!\!\sf}_{\!L}}_{jimm} - Y^f_L Y^{\hat{f}}_R \,R^{\sf\
\!\hat{\!\!\sf}_{\!D}}_{jimm} - Y^{\hat{f}}_L Y^f_R \,R^{\
\hat{\!\!\sf}\sf_D}_{mmji}
\\[1mm]
&&\hspace{70mm} +~Y^f_R Y^{\hat{f}}_R R^{\sf\
\!\hat{\!\!\sf}_{\!R}}_{jimm} \bigg] A_0(m_{\
\!\hat{\!\!\sf}_{\!m}}^2)
\\[2mm]\non
\Pi_{ij}^{\sf,\, \ \!\hat{\!\!\sf}\,\!'} &=&  -\frac{1}{(4\pi)^2} 
N_C^{\hat f}\, \frac{g^2}{4} \sum_{m=1}^2 R^{\sf\ 
\!\hat{\!\!\sf}\,\!'_{\!L}}_{jimm} A_0(m_{\ 
\!\hat{\!\!\sf}\,\!'_{\!m}}^2) 
\\[1mm] \non
&&+\frac{1}{(4\pi)^2}\, N_C^{\hat f}\, \frac{g'^2}{4} \sum_{m=1}^2
\bigg[ Y^f_L Y^{\hat{f}'}_L R^{\sf\
\!\hat{\!\!\sf}\,\!'_{\!L}}_{jimm} - Y^f_L Y^{\hat{f}'}_R
\,R^{\sf\ \!\hat{\!\!\sf}\,\!'_{\!D}}_{jimm} - Y^{\hat{f}'}_L
Y^f_R \,R^{\ \hat{\!\!\sf}\,\!'\!\sf_D}_{mmji}
\\[1mm]
&&\hspace{70mm} +~Y^f_R Y^{\hat{f}'}_R R^{\sf\
\!\hat{\!\!\sf}\,\!'_{\!R}}_{jimm} \bigg] A_0(m_{\
\!\hat{\!\!\sf}\,\!'_{\!m}}^2)
\end{eqnarray}
The contributions from first and second generation sfermions, $\ti 
F_m$, are given by 
\begin{eqnarray}
\begin{array}{r@{\ }c@{\ }l@{\qquad\quad}c@{\ }r@{\ }l}
\Pi_{ij}^{\sf,\, \ti F} &=& \Pi_{ij}^{\sf,\, \ \!\hat{\!\!\sf}} 
(\hat f \rightarrow F)\,, & \Pi_{ij}^{\sf,\ \!\hat{\!\ti F}} &=& 
\Pi_{ij}^{\sf,\, \ \!\hat{\!\!\sf}} (\hat f \rightarrow \hat F)\,,
\\[3mm]
\Pi_{ij}^{\sf,\, \ti F'} &=& \Pi_{ij}^{\sf,\, \ 
\!\hat{\!\!\sf}\,\!'} (\hat f' \rightarrow F')\,, & 
\Pi_{ij}^{\sf,\ \hat{\!\ti F}\,\!'} &=& \Pi_{ij}^{\sf,\, \ 
\!\hat{\!\!\sf}\,\!'} (\hat f' \rightarrow \hat F')\,,
\end{array}
\end{eqnarray}
where the sub--/superscript $\ti F$ denotes values belonging to 
first and second generation scalar fermions with same isospin as 
$\sf$ (e.~g. $\ti F_1 = \{ \ti u_1, \ti c_1 \}$ for the stop case, 
\ldots), $\ti F'$ sfermions with different isospin etc. 
\begin{eqnarray}\non
\Pi_{ij}^{\sf,\,H} &=& \frac{1}{(4\pi)^2} \,\frac{1}{2} 
\sum_{k=1}^4 \bigg[ h_f^2 \,c^\sf_{kk}\, \d_{ij} + g^2 
\,e^\sf_{ij}\, \Big( c^\sb_{kk} - c^\st_{kk} \Big) \bigg] A_0 
(m_{H_k^0}^2) 
\\[1mm] \non
&&\hspace{-3mm}+\frac{1}{(4\pi)^2} \sum_{k=1}^2 \bigg[ h_{f'}^2
\,d^{\sf'}_{kk}\, R^\sf_{i1} R^\sf_{j1} + h_{f}^2 \,d^{\sf}_{kk}\,
R^\sf_{i2} R^\sf_{j2} + g^2 \,f^\sf_{ij}\, \Big( d^\sb_{kk} -
d^\st_{kk} \Big) \bigg] A_0 (m_{H_k^+}^2)
\\
\\ \non
\Pi_{ij}^{\sf,\,V} &=& \frac{1}{(4\pi)^2} \,4\, (e_0 e_f)^2 
\d_{ij} A_0(\l^2) + \frac{1}{(4\pi)^2} \,2\, g^2 R^\sf_{i1} 
R^\sf_{j1} A_0(m_{\scriptscriptstyle W}^2) 
\\[1mm]
&&\hspace{-3mm}+\frac{1}{(4\pi)^2} \,4\, g_{\scriptscriptstyle 
Z}^2 \bigg[ \Big( C_L^f \Big)^2 R^\sf_{i1} R^\sf_{j1} + \Big( 
C_R^f \Big)^2 R^\sf_{i2} R^\sf_{j2} \bigg] 
A_0(m_{\scriptscriptstyle Z}^2) 
\end{eqnarray}

\begin{figure}[th]
\begin{picture}(170,105)(0,0)
     \put(0,0){\mbox{\resizebox{16cm}{!}
     {\includegraphics{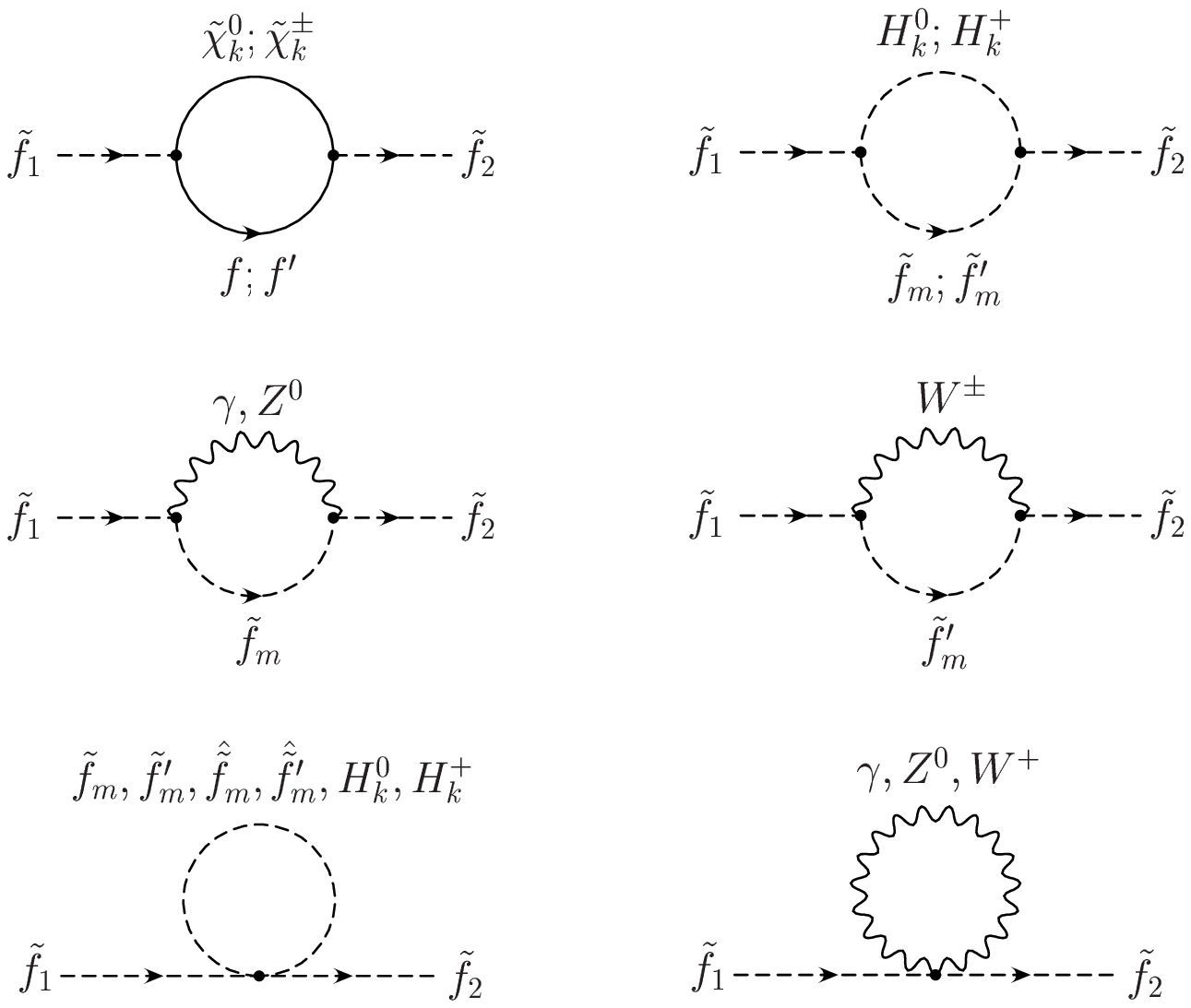}}}}
\end{picture}
\caption{Off--diagonal Sfermion self--energies\label{SfermionSEoffdiag}}
\end{figure}

\end{document}